\documentstyle[pra,aps,eqsecnum,twoside,multicol,rotate,epsf]{revtex}

\begin{document}
\draft

\pagestyle{myheadings}
\markboth{D.A. Lidar \& O. Biham }{Phys. Rev. E. {\bf 56}, 3661 (1997)}

\newcommand{\be}{\begin{equation}}
\newcommand{\ee}{\end{equation}}
\newcommand{\ba}{\begin{eqnarray}}
\newcommand{\ea}{\end{eqnarray}}
\newcommand{\ra}{\rangle}
\newcommand{\la}{\langle}
\newcommand{\yb}{\overline{y}}

\title{Simulating Ising Spin Glasses on a Quantum Computer}

\author{Daniel A. Lidar$^{a,b}$\footnote{Formerly: Hamburger. URL:
http://neon.cchem.berkeley.edu/$\sim$dani. Email:
dani@neon.cchem.berkeley.edu. Current address:
Chemistry Department, University of California, Berkeley, CA 94720  
}
and Ofer Biham$^{a}$\footnote{URL:
http://www.fiz.huji.ac.il/staff/acc/faculty/biham. Email:
biham@flounder.fiz.huji.ac.il}}
\address{
$^{a}$
Racah Institute of Physics, The Hebrew University, Jerusalem 91904, Israel
}
\address{
$^{b}$
The Fritz Haber Center for Molecular Dynamics, The Hebrew University,
Jerusalem 91904, Israel
}

\maketitle

\date{\today}

\begin{abstract}

A linear-time algorithm is presented for the construction of the
Gibbs distribution of configurations in the Ising model, on a quantum
computer. The algorithm is designed so that each run provides one
configuration with a quantum probability equal to the corresponding
thermodynamic weight. The partition function is thus approximated 
efficiently. The algorithm neither suffers from critical
slowing down, nor gets stuck in local minima. 
The algorithm can be applied in
any dimension, to a class of spin-glass Ising models with a finite
portion of frustrated plaquettes, diluted Ising models, and models
with a magnetic field.

\end{abstract}

\begin{multicols}{2}

\section{Introduction}
\label{Introduction}

The algorithm of Shor \cite{Shor:94} for the polynomial time solution
of the factorization problem on a quantum computer was received with
much excitement in the computer science and physics communities
\cite{Barenco:2,Ekert96}. It indicates that quantum
computers have a potential for the effective solution of problems
which are unfeasible on a classical computer.  The actual utilization
of this potential would require, in addition to many years of work on
the hardware, the development of algorithms which would optimally
exploit the strengths while overcoming the shortcomings of quantum
computers, in particular the problem of decoherence
\cite{DiVincenzo,Zurek}. Unlike classical computers in which each bit
is a two state system which can be either in state 0 or 1 the quantum
bit (or qubit) can be in any superposition of the form:

\begin{equation}
|\psi \rangle = \alpha_0 |0 \rangle + \alpha_1 |1 \rangle ,
\label{qubit}
\end{equation}

\noindent as long as $|\alpha_0|^2 + |\alpha_1|^2 =1$. When a
measurement of the qubit takes place the result will be the state $|0
\rangle$ with probability $|\alpha_0|^2$ or $|1
\rangle$ with probability $|\alpha_1|^2 $. In the first case the
system will then remain in the state $|0 \rangle$ while in the second
case it will remain in the state $|1 \rangle$. Due to superposition, a
system of $N$ qubits is described by a unit vector in a $2^N$
dimensional complex vector space (the Hilbert space) of the form:

\begin{equation}
|\psi_N \rangle = \sum_{i=0}^{2^N -1} \alpha_i |i\rangle ,
\end{equation}

\noindent where $|i\rangle$ are the $2^N$ basis vectors and $\sum
|\alpha_i|^2 =1$. Computations are done by changing the state of the
system.  Since conservation of probability is required only unitary
transformations are allowed. One source of the potential strength of quantum
computers is due to the fact that computations are performed
simultaneously on all $2^N$ states in the superposition. This amounts
to an exponential parallelism compared to classical computers in which
at any given time only one number can be stored in the $N$ bit
register. However, this feature alone is not enough since only a
single quantum state is available after measurement
\cite{QC:comment11}. The full power of quantum computation is realized
only when the superposition principle is combined with the ability of
quantum states to {\em interfere}. The latter has no classical
analogue, and is the quantum ingredient which allows one to
selectively control which
state will have the highest probability of appearing after
measurement. 

Similarly to classical computers, it turns out that all unitary
transformations involving $N$ qubits can be broken into two qubit
unitary transformations
\cite{DiVincenzo95,Sleator:95,Lloyd:95,Deutsch95}. This allows one to
construct a universal set of binary gates which is capable of
implementing all the required operations.  The actual construction of
a quantum computer is a formidable task that is believed to require
many years before a basic prototype will be ready.  Some of the
potential physical media proposed for quantum computers include ions
in ion traps \cite{Cirac95}, atoms coupled to optical resonators
\cite{Turchette}, interacting electrons in quantum dots
\cite{Barenco95}, and Ramsey atomic interferometry \cite{Brune94}.
 
The main difficulty identified so far in the construction of a quantum
computer is the decoherence of the quantum superposition due to
interaction with the environment. To avoid errors one needs to isolate
the quantum computer from the environment as much as possible.  Some
redundancy combined with error correcting codes is considered as a
promising way to reduce the accumulation of errors during the
computation
\cite{Shor95,Steane,Calderbank,Ekert,Laflamme,Bennett:1,Aharonov:1,Peres}.
Another potential difficulty is to maintain sufficient precision so
that the quantum computer will provide accurate results even after
many steps of computing. Therefore, an efficient quantum algorithm
should satisfy not only that the time and memory required for the
computation grow polynomially with the input size, but also the
precision: the number of bits of precision should grow only
logarithmically in the input size \cite{Shor:94,Bennett:2}.

In addition to the prospects of building a quantum computer, the
experimental work stimulated by this field is expected to provide new
insights on the foundations of quantum mechanics, as well as to lead
to progress in the development of new technologies. Furthermore, the
new perspective that quantum computers provide about the complexity of
algorithms is highly valuable for the theory of computation even
without the physical construction of such a computer.  In particular,
Shor's algorithm for the polynomial time solution of the
factorization problem \cite{Shor:94} has shown that problems which are
considered intractable on classical computers may be tractable on a
quantum computer, although some restrictions are known related to the
potential power of a quantum computer \cite{Bennett:2}. Feynman
\cite{Feynman:QC} was the first to suggest that quantum computers
might be exponentially faster than classical computers at simulating quantum
mechanical systems with short range interactions. A general demonstration to
this effect was given by Lloyd \cite{Lloyd:96}, who also
argued that quantum computers could efficiently calculate spin-spin
correlation functions in Ising models. Some explicit algorithms were
later proposed for simulating physical systems on quantum
computers. These include the Schr\"{o}dinger equation for interacting
many-body systems \cite{Boghosian:1,Boghosian:2,Wiesner,Zalka}, the
Dirac equation \cite{Meyer}
and the quantum baker's map \cite{Schak97}.

In this paper we consider a broad class of statistical physics
problems involving Ising spin systems. We develop an algorithm for
simulating these systems on a quantum computer.
Here we define the Ising spin systems and briefly review the numerical
techniques in use for their simulation on classical computers. These
systems are important both as models of magnetic phase transitions and
as the most useful models for the analytical and numerical studies of
phase transitions in general \cite{Stanley:book,Binney:book}. The
numerical simulation of such systems has been an active field of
research for the last five decades since the introduction of the
Metropolis algorithm \cite{Metropolis}. Typically, spin systems are
defined on a $d$ dimensional lattice in which there are $N$ spins
$s_i$, one 
at each lattice site $i$, and nearest neighbor coupling between spins.
The energy of the system is given by the nearest-neighbor
Edwards-Anderson Hamiltonian \cite{Edwards}:

\begin{equation}
H = - \sum_{\langle i,j \rangle} J_{ij} s_i s_j - \sum_{i=1}^N h_i s_i
,
\end{equation}

\noindent where 
$\langle i,j \rangle$ represents summation only over pairs of nearest
neighbors, $J_{ij}$ is the coupling between $i$ and $j$, and $h_i$ is
the local magnetic field.  The most
commonly studied model is the Ising model \cite{Ising} in which each
spin has two states $s_i = \pm 1$.  The bonds $J_{ij}$ then determine
the nature of the interactions.  In the ordinary ferromagnetic
(antiferromagnetic) Ising model all bonds satisfy $J_{ij} \! = \! J$
($J_{ij} \! = \! - J$) where $J>0$.  In the $\pm J$ Ising spin glass there
are quenched random bonds chosen from a bimodal distribution
$P(J_{ij}) \! = \! p \delta (J_{ij} - J) + (1-p) \delta (J_{ij} + J)$.  The
random bonds in the Ising spin glass can also be drawn from a
continuous distribution such as the Gaussian distribution.

Numerical studies of spin systems have been performed for a vast
variety of lattices including the square and triangular lattices in
two dimensions (2D), cubic, hexagonal and hexagonal closed packed in
three dimensions (3D). Here we will concentrate mainly on finite
hypercubic lattices in $d$ dimensions, which include $N\! =\! L^d$ sites,
where $L$ is the number of sites along each edge.  Since each spin has
two states, these Ising spin systems exhibit an exponentially large
phase space of $2^N$ configurations.  The partition function of Ising
type models is given by a sum over all configurations: $Z =
\sum_{\{s\}} \exp(-\beta H[\{s\}])$ where $\beta \! = \! 1/(k_B T)$, $k_B$
is the Boltzmann constant and $T$ is the temperature.  For a system in
thermodynamic equilibrium, the partition function provides the
statistical weight of each one of the $2^N$ configurations.  The
statistical weight of the configuration $\{s_i\}, i=1,\dots,N$ is
given by: ${\rm Pr}[\{s_i\}] \! = \! \exp(-\beta H[\{s_i\}])/Z$.  Therefore,
if the partition function is known one can obtain exact results for
all the thermodynamic quantities such as the magnetization,
susceptibility and specific heat.  Models for which analytical
calculations of this type can be performed include a variety of one
dimensional (1D) models and the 2D Ising model \cite{Onsager44}.
However, for most systems of interest, including the $3D$ Ising model
and most Ising spin glass models, no analytical calculation of the
partition function is available \cite{Green}.

The size of the input in computations of Ising spin systems is simply
the number of spins $N$ plus the number of bonds $N_B$ which connect
these spins.  In models of short range interaction considered here the
number of bonds is of order $O(N)$. An exact numerical calculation of
the partition function or any thermodynamic quantity would involve a
summation over the $2^N$ terms which appear in the partition function.
As the system size increases the computation time would increase
exponentially, and this is obviously not feasible. In order to obtain
thermodynamic averages a variety of Monte Carlo methods have been
developed. These methods involve a sequential random sampling of the
phase space moving from one configuration to the next according to a
properly designed Markov process. In order to sample all
configurations with the appropriate thermodynamic weights, the Markov
process must be able to access the entire phase space and to satisfy
the detailed balance condition:

\begin{equation}
{\rm Pr}(\{s_i\}) \cdot W(\{s_i\} \rightarrow \{s_i\}^{\prime}) =
{\rm Pr}(\{s_i\}^{\prime}) \cdot W(\{s_i\}^{\prime} \rightarrow \{s_i\})
\end{equation}

\noindent where 
$\{s_i\}$ and $\{s_i\}^{\prime}$ are any two states of the system and
$W(\{s_i\} \rightarrow \{s_i\}^{\prime})$ is the transition
probability from one state to the other in a single move of the Markov
process \cite{Binney:book}. When these conditions are satisfied one
can use the Monte Carlo results to obtain approximations to
thermodynamic quantities.

In the most commonly used Metropolis algorithm \cite{Metropolis}, at
each time step one spin is chosen randomly. The energy difference
$\Delta E$ which would occur due to flipping the chosen spin is
calculated. If $\Delta E \le 0$ the move is accepted and the spin is
flipped. If $\Delta E > 0$ the move is accepted with probability $p =
e^{-\beta \Delta E}$. Since this rule satisfies detailed balance, one
can take samples of the configurations during the run to obtain
properties of the equilibrium phases such as magnetization,
susceptibility, correlation function, correlation length, etc.
Although for large systems it is feasible to scan only an
exponentially small part of the phase space, this part typically has
an exponentially large weight and therefore, Monte Carlo simulations
provide good approximations for the thermodynamic quantities.

The Metropolis algorithm and related techniques which involve flipping
one spin at a time are efficient as long as the system is not too
close to a critical point, i.e., a second order phase transition.
Near the critical point the simulation suffers from ``critical slowing
down'' and the number of Monte Carlo steps needed between uncorrelated
configurations grows as $L^{z}$ where $z$ is the dynamical critical
exponent \cite{Binney:book}. The reason for this slow dynamics is
that near the critical point there are large clusters of highly
correlated spins. In this situation there is a very small likelihood
of flipping an entire cluster due to the large energy barrier
involved. To overcome this difficulty, cluster algorithms were
introduced, in which entire clusters are flipped at once, in a way
that maintains detailed balance \cite{Swendsen87}.

In addition to the regular lattice spin systems, there has been much
interest in the study of disordered systems such as frustrated Ising
models \cite{Toulouse,Kirkpatrick77} and the Ising spin glass
\cite{Parisi:book,Binder86}. These systems exhibit competing
ferromagnetic and antiferromagnetic interactions. In particular, in
plaquettes which include an odd number of antiferromagnetic bonds it
is not possible to satisfy all the bonds simultaneously and the system
is thus frustrated \cite{Toulouse,Kirkpatrick77}. Spin glasses
exhibit a complex energy landscape with a large number of metastable
states or local minima. Since these minima are separated by energy
barriers, when the system is simulated using Monte Carlo methods at
low temperatures, it tends to be trapped around one local minimum from
which it cannot escape. When the simulation is done at high
temperature, the system can easily switch from the vicinity of one
local minimum to another but cannot resolve the details, namely reach
the local minimum itself. The approach of simulated annealing
\cite{Kirkpatrick:83} in which the temperature is repeatedly raised
and then slowly decreased was found useful for obtaining thermodynamic
averages. In particular, it provides a probabilistic algorithm for
exploring the local minima and for searching for the ground state of the
system. The problem of finding the ground state of the short range 3D
Ising spin glass, as well as the fully antiferromagnetic 2D Ising
model in the presence of a constant magnetic field was shown by
Barahona to belong to the
class of 
non-deterministic polynomial time
(NP)-hard problems, by a mapping to problems in graph
theory \cite{Barahona,Parisi86,Fu86,Anderson86}.

In this paper we present an algorithm for the
study of a class of random-bond Ising spin systems on a quantum
computer. By use of interference, the algorithm can construct, with linear complexity, a lattice with a fixed portion of
plaquettes having predetermined bonds. The bonds connecting plaquettes are
determined 
randomly, with the probability of obtaining a non-frustrated intermediate
plaquette being higher than that of a frustrated one. The
superposition property of a quantum computer can be used in order to
include the 
entire phase space of the resulting Ising system, such that the quantum
mechanical probability of each one of the $2^N$ states equals the
thermodynamic weight of the corresponding spin configuration. In this
sense the algorithm is exact. Once
such a superposition is constructed one can perform a measurement of
all spins, which provides one of the $2^N$ configurations. Since the
probabilities of the quantum states are ordered by the thermodynamic
weights of the corresponding spin configurations, the partition function is
constructed efficiently.
Putting aside questions of degeneracy, the lower
the energy of the configuration, the more likely it is to be obtained
upon measurement. Therefore the algorithm can be used to find
ground state configurations of the spin system.
Unlike Monte Carlo
simulations which require a minimal number of steps 
between measurements to reduce the
autocorrelation function between consecutively measured configurations
\cite{Binney:book}, measurements on the quantum computer are totally
uncorrelated since the superposition is constructed anew before every
measurement. Successive runs of the algorithm
therefore involve no dynamics of moving from one configuration to
another. A situation in which this procedure is especially useful, is in
the vicinity of a critical point, where Monte Carlo simulations may suffer
from critical slowing down. While cluster algorithms have been
invented for regular Ising models \cite{Swendsen87,Kandel90} which
essentially solve this problem, they are very
limited in scope and can treat essentially only Ising systems with
periodic bond-structure. The present algorithm is more general in
that it applies to random-bond Ising models as well, and avoids the
problem of critical slowing down altogether.

The paper is organized as follows. The construction of the
superposition of states for the 1D Ising model, with quantum
probabilities equal to the thermodynamic weights is considered in
Section ~\ref{1D}. Higher dimensional Ising systems including the
Bethe lattice, the 2D and 3D Ising models are considered in Section
~\ref{highD}. A magnetic field is introduced in Sec.\ref{field}. The
conclusions appear in Section ~\ref{conclusions}.

\section{1D Ising Model}
\label{1D}

We begin our exposition of the algorithm by treating the simple case
of a 1D Ising model. Starting from the fully ferromagnetic open chain,
we will gradually introduce complexity, by considering the
antiferromagnetic case, mixed ferro-antiferromagnetic models,
spin-glasses, and finally close the boundary conditions. This last
operation, which enables the use of transfer matrices in the classical
case, allows for a comprehensive treatment of the 2D and higher
dimensional models.

\subsection{The Ferromagnetic Case}
\label{ferro}

The Hamiltonian for a linear, open ferromagnetic
system of $N$ spins $s_i = \pm 1$ is \cite{QC:comment1}:

\be
H^+_N = -J \sum_{i=1}^{N-1} s_i s_{i+1} ,
\label{eq:H_N}
\ee

\noindent where $J>0$. Let $y \in
[0,2^N-1]$, and $\{s\}_y^N$ be the $N$-digit binary expansion of $y$ using
$s_i=-1$ for 0 and $s_i=+1$ for 1. The notation $\{s\}_y^N$ can also
denote one of the $2^{N}$ spin configurations, with thermodynamic weight:

\be
{\rm Pr}[\{s\}_y] = {1 \over Z^+_N} e^{-\beta H^+_N[\{s\}_y]} ,
\label{eq:Pr-1D}
\ee

\noindent (the $N$ superscript on $\{s\}_y^N$ will be suppressed where
it is obvious) where:

\be
Z^+_N = \sum_{y=0}^{2^N-1} e^{-\beta H^+_N[\{s\}_y]} ,
\ee

\noindent is the partition
function. It is the task of the algorithm to exactly calculate the
probabilities above, in a manner which allows an easy identification
of the configuration whose probability was found. To this end we
introduce an $N$-qubit register $\{s\}_y^N = | s_1, s_2,...,s_N \ra$,
where now $s_i = \pm 1$ denote the ground and excited states of the
$i^{\rm th}$ qubit (it is convenient to use the same notation for the classical
spins and the qubits, and this should not cause any confusion). We
term this a register of ``spin-qubits''. Let
$|\{-\}_N\ra$ denote the ground state of all spin-qubits. We seek a unitary
operator $T^+_N$ which simulates the Ising model in the following sense:

\be
\left| \la \{s\}_y^N | T^+_N |\{-\}_N\ra \right|^2 = {\rm Pr}[\{s\}_y] .
\label{eq:sol1D}
\ee

\noindent Thus $T^+_N$ evolves the qubit register into a superposition
in which every state uniquely codes for an Ising configuration of
spins, with a quantum probability equal to the thermodynamic weight of
that configuration. $T^+_N$ must be a ``valid'' quantum computer
operator, i.e., it must be decomposable into a product of a polynomial
(in $N$) number of 1- and 2-qubit unitary operators only
\cite{DiVincenzo}. Such a decomposition is possible with the following
two operators: a 1-qubit $\pi/2$ rotation,

\ba
\lefteqn{ R_i | s_1,...,s_i,...,s_N \ra = } \nonumber \\
&&
{1\over \sqrt{2}} (| s_1,...,-s_i,...,s_N \ra
- s_i | s_1,...,s_i,...,s_N \ra) ,
\label{eq:R}
\ea

\noindent and a 2-qubit ``Ising-entanglement'':

\ba
\lefteqn{ S^+_{ij} | s_1,...,s_i,...,s_j,...,s_N \ra = } \nonumber \\
&&
{1\over \sqrt{c}} \left( x^{-J s_i} | s_1,...,s_i,...,s_j,...,s_N
\ra + \right. \nonumber \\
&&
\left. s_j x^{J s_i} | s_1,...,s_i,...,-s_j,...,s_N \ra \right) ,
\label{eq:S}
\ea

\noindent where

\ba
x \equiv e^{\beta/2}, \:\:\:\: c \equiv {1\over 2} Z^+_2 = 2\cosh(\beta J) .
\label{eq:x,c}
\ea

\noindent In what follows we will suppress the full register and
indicate only the qubits operated on. It is straightforward to check
that $R_i$ and $S^+_{ij}$ are indeed unitary, e.g., by considering their
matrix representations in the basis where $|-\ra = (1,0)$, $|+\ra =
(0,1)$, $|--\ra = (1,0,0,0)$, $|-+\ra = (0,1,0,0)$, $|+-\ra =
(0,0,1,0)$, $|++\ra = (0,0,0,1)$:

\ba
R = {1\over \sqrt{2}}
\left(
\begin{array}{rr}
1 &  1 \\
1 & -1
\end{array}
\right)
\label{eq:R-mat}
\ea

\noindent and

\be
S^+ = {1 \over \sqrt{c}}
\left(
\begin{array}{clll}
 x^J    & x^{-J} & 0              & 0      \\
-x^{-J} & x^J    & 0              & 0      \\
 0      & 0      & x^{-J}         & x^J    \\
 0      & 0      & \!\!\!\!\!-x^J & x^{-J} \\
\end{array}
\right) .
\label{eq:S-mat}
\ee

\noindent It is interesting to note the similarity to the classical 1D
transfer matrix,

\be
\left(
\begin{array}{cl}
 x^{2J}    & x^{-2J} \\
-x^{-2J} & x^{2J}
\end{array}
\right) .
\label{eq:transfer-mat}
\ee

\noindent The $x^J$ {\it vs} $x^{2J}$ comes from the fact that in $S^+$ we
have amplitudes, not probabilities.

The operator which simulates the 1D Ising problem can now be
written as:

\be
T^+_N = \left[ \prod_{i=N-1}^{1} S^+_{i,i+1} \right] R_1 .
\label{eq:T^+_N}
\ee

\noindent Thus $T^+_N$ is a $\pi/2$ rotation of the first qubit,
followed by Ising entanglements of successive pairs of qubits. This
bares some resemblance to the procedure using a transfer matrix to
solve the 1D Ising model. The
number of required operations is exactly $N$. The general
``recipe'' for writing down this operator (in the absence of
closed loops) is the following: one always
applies a $\pi/2$ rotation to the first qubit, and then substitutes an
Ising-entanglement operator for each interacting pair of
nearest-neighbor spins in the Hamiltonian. 

\end{multicols}
%%\hrulefill\

\begin{table}
\begin{tabular}{|l|c|c|c|c|c|} \hline
1 & 2 & 3 & 4 & 5 & 6 \\ \hline
$\left|\right.s_1, s_2, s_3, s_4 \ra$ & initial & $R_1$ & $S^+_{12}$ &
$S^+_{23}$ & $S^+_{34}$ \\ \hline \hline
$\left|\right.----\rangle$& 1 & $1/\sqrt{2}$ & $x^J/\sqrt{2c}$    &
$x^{2J}/\sqrt{2}c$ & $x^{3J}/\sqrt{2}c^{3/2}$      \\ \hline
$\left|\right.---+\rangle$& 0 & 0          & 0                   &
0     & $-x^J/\sqrt{2}c^{3/2}$             \\ \hline
$\left|\right.--+-\rangle$& 0 & 0          & 0                   &
$-1/\sqrt{2}c$   & $-x^{-J}/\sqrt{2}c^{3/2}$       \\ \hline
$\left|\right.--++\rangle$& 0 & 0          & 0                   &
0     & $x^J/\sqrt{2}c^{3/2}$              \\ \hline
$\left|\right.-+--\rangle$& 0 & 0          & $-x^{-J}/\sqrt{2c}$  &
$-x^{-2J}/\sqrt{2}c$  & $-x^{-J}/\sqrt{2}c^{3/2}$  \\ \hline
$\left|\right.-+-+\rangle$& 0 & 0          & 0                   &
0         & $x^{-3J}/\sqrt{2}c^{3/2}$      \\ \hline
$\left|\right.-++-\rangle$& 0 & 0          & 0                   &
$1/\sqrt{2}c$       & $x^{-J}/\sqrt{2}c^{3/2}$     \\ \hline
$\left|\right.-+++\rangle$& 0 & 0          & 0                   &
0         & $-x^J/\sqrt{2}c^{3/2}$         \\ \hline
$\left|\right.+---\rangle$& 0 & $1/\sqrt{2}$ & $x^{-J}/\sqrt{2c}$ &
$1/\sqrt{2}c$       & $x^J/\sqrt{2}c^{3/2}$        \\ \hline
$\left|\right.+--+\rangle$& 0 & 0          & 0                   &
0         & $-x^{-J}/\sqrt{2}c^{3/2}$      \\ \hline
$\left|\right.+-+-\rangle$& 0 & 0          & 0                   &
$-x^{-2J}/\sqrt{2}c$  & $-x^{-3J}/\sqrt{2}c^{3/2}$ \\ \hline
$\left|\right.+-++\rangle$& 0 & 0          & 0                   &
0         & $x^{-J}/\sqrt{2}c^{3/2}$       \\ \hline
$\left|\right.++--\rangle$& 0 & 0          & $-x^J/\sqrt{2c}$     &
$-1/\sqrt{2}c$       & $-x^J/\sqrt{2}c^{3/2}$      \\ \hline
$\left|\right.++-+\rangle$& 0 & 0          & 0                   &
0         & $x^{-J}/\sqrt{2}c^{3/2}$       \\ \hline
$\left|\right.+++-\rangle$& 0 & 0          & 0                   &
$x^{2J}/\sqrt{2}c$    & $x^J/\sqrt{2}c^{3/2}$      \\ \hline
$\left|\right.++++\rangle$& 0 & 0          & 0                   &
0         & $-x^{3J}/\sqrt{2}c^{3/2}$      \\ \hline
\end{tabular}
\caption{Amplitudes of register states for 1D Ising model with up to 4 spins.}
\label{tab:ex1D}
\end{table}

%%\hrulefill\
\begin{multicols}{2}

\noindent It might be helpful to give an example at this point. For an open
chain of $N \leq 4$ spins, Table \ref{tab:ex1D} gives the amplitudes
of four spins, at each stage of the algorithm, as calculated from
Eq.(\ref{eq:T^+_N}). It is easily verified that the squares of the
amplitudes given in columns 4, 5, 6, agree with the thermodynamic
weights [given by Eq.(\ref{eq:H_N})] for $N$=2, 3, 4 spins
respectively, with a ferromagnetic interaction. (In order to check,
e.g, for $N$=2, ignore the entries for $s_3$ and $s_4$.)

We proceed to prove that $T^+_N$ indeed satisfies
Eq.(\ref{eq:sol1D}), by mathematical induction. Let us consider first
the minimal case $N=2$ (for $N=1$ we cannot apply $S^+$). We have
$T^+_2 = S^+_{12} R_1$, so that

\ba
\lefteqn{ T^+_2 |\{-\}_2\ra = {1 \over \sqrt{2}} S^+_{12} (|--\ra + |+-\ra)
 = } \nonumber \\
&&
{1 \over \sqrt{2c}} \left[(x^J|--\ra - x^{-J}|-+\ra) + (x^{-J}|+-\ra -
x^J |++\ra)\right] = \nonumber \\
&&
{1 \over \sqrt{2c}} \sum_{s_1,s_2=\pm 1} -s_2 x^{J s_1 s_2} |s_1,s_2\ra .
\ea

\noindent On the other
hand, according to Eq.(\ref{eq:Pr-1D}) the thermodynamic weights of
these four states are, respectively, $e^{\beta J}/Z^+_2$, $e^{-\beta J}/Z^+_2$,
$e^{-\beta J}/Z^+_2$, $e^{\beta J}/Z^+_2$. These are exactly the squares of the
above amplitudes.
Next, assume by induction that Eq.(\ref{eq:sol1D}) holds for $T^+_N$,
i.e., that

\ba
T^+_N |\{-\}_N\ra = \sum_{y=0}^{2^N-1} A^+_{y} |\{s\}_y \ra ,
\ea

\noindent where:
\ba
\lefteqn{
A^+_{y} \equiv {1 \over \sqrt{Z^+_N}} x^{-H^+_N[\{s\}_y]} \phi_N}
\nonumber \\
&&
\phi_N \equiv -\prod_{i=2}^N (-s_i) .
\label{eq:phi_N}
\ea

\noindent Using this, we must show that $T^+_{N+1}$ satisfies
Eq.(\ref{eq:sol1D}) for the 1D Ising model with $N+1$ spins. Now,

\ba
\lefteqn{ T^+_{N+1} |\{-\}_{N+1}\ra = S^+_{N,N+1} T^+_N |\{-\}_N\ra |-\ra = 
} \nonumber \\
&&
\sum_{y=0}^{2^N-1} A^+_{y} S^+_{N,N+1} |\{s\}_y^N \ra |-\ra ,
\label{eq:T^+_N+1}
\ea

\noindent where the
last equality follows from the induction hypothesis. But by definition
of $S^+$,

\ba
S^+_{N,N+1} |\{s\}_y^N\ra |-\ra = {1 \over \sqrt{c}}
\left( x^{-J s_N} |\{s\}_y^N\ra |-\ra - x^{J s_N} |\{s\}_y^N\ra
|+\ra \right) . \nonumber
\ea

\noindent Inserting this into Eq.(\ref{eq:T^+_N+1}), we have:

\ba
\lefteqn{ T^+_{N+1} |\{-\}_{N+1}\ra = } \nonumber \\
&&
\sum_{y=0}^{2^N-1} {\phi_N \over \sqrt{Z^+_N}} e^{-\beta H^+_N[\{s\}_y]/2}
{1 \over \sqrt{c}} \times \nonumber \\
&&
\left( e^{-s_N \beta J/2} |\{s\}_y^N,-\ra - e^{s_N \beta J/2}
|\{s\}_y^N,+\ra \right) .
\label{eq:T^+_N+1:2}
\ea

\noindent The
two terms in this sum arise from the two possible states of
$|s_{N+1}\ra$, so writing the exponents as $\exp(s_N s_{N+1} \beta J/2)$
and using Eq.(\ref{eq:H_N}), we obtain:

\ba
T^+_{N+1} |\{-\}_{N+1}\ra =
\sum_{y=0}^{2^{N+1}-1} {\phi_{N+1} \over \sqrt{Z^+_{N+1}}}  e^{-\beta
H_{N+1}[\{s\}_y]/2} |\{s\}_y^{N+1}\ra . \nonumber
\ea

\noindent The state $|\{s\}_y^{N+1}\ra$ thus
appears with a probability equal to its thermodynamic weight, which
completes our proof. A useful corollary is that

\newtheorem{cor0}{Corollary}
\begin{cor0}
Upon a measurement following the execution of the algorithm, a state
appears with a probability equal to the thermodynamic weight of the
corresponding spin configuration.
\label{cor0}
\end{cor0}

This implies that the present algorithm provides an approximation to
the partition function which converges rapidly in the number of
measurements.

\subsection{The Antiferromagnetic Case}
\label{antiferro}

In the antiferromagnetic case, the Hamiltonian for a linear, open 
system of $N$ spins is

\be
H^-_N = J \sum_{i=1}^{N-1} s_i s_{i+1} ,
\label{eq:H^-_N}
\ee

\noindent where $J>0$. To treat this case we define a properly
modified version of the Ising-entanglement operator $S^+$:

\be
S^-_{ij} | s_i,s_j \ra = {1 \over \sqrt{c}} \left( x^{J s_i} | s_i,s_j \ra +
s_j x^{-J s_i} | s_i,-s_j \ra \right) .
\label{eq:S^-}
\ee

\noindent In matrix form, 

\be
S^- = {1 \over \sqrt{c}}
\left(
\begin{array}{llcl}
 x^{-J}        & x^J    & 0       & 0      \\
\!\!\!\!\!-x^J & x^{-J} & 0       & 0      \\
 0             & 0      & x^J     & x^{-J} \\
 0             & 0      & -x^{-J} & x^J    \\
\end{array}
\right)
\label{eq:S^--mat}
\ee

\noindent and it is easily checked that $S^-_{ij}$ is unitary. We claim
that

\be
T^-_N = \left[ \prod_{i=N-1}^{1} S^-_{i,i+1} \right] R_1
\label{eq:T^-_N}
\ee

\noindent simulates the antiferromagnetic ``1D Ising model'', in the sense of
Eq.(\ref{eq:sol1D}). The proof is entirely analogous to the
ferromagnetic case. First one considers the case $N=2$, for which:

\ba
\lefteqn{ T^-_2 |\{-\}_2\ra = 
{1 \over \sqrt{2}} S^-_{12} (|--\ra + |+-\ra) = } \nonumber \\
&&
{1 \over \sqrt{2c}} \left[(x^{-J}|--\ra - x^J |-+\ra) + (x^J |+-\ra -
x^{-J}|++\ra)\right] . \nonumber
\ea

\noindent The state-probabilities are just the thermodynamic weights
which can be obtained from Eq.(\ref{eq:H^-_N}) for $N=2$. Repeating
the induction argument that led to Eq.(\ref{eq:T^+_N+1:2}) one finds
here:

\ba
\lefteqn{ T^-_{N+1} |\{-\}_{N+1}\ra = } \nonumber \\
&&
\sum_{y=0}^{2^N-1} {\phi_{N} \over \sqrt{Z^-_N}} x^{-H^-_N[\{s\}_y]}
\sqrt{1 \over c} \times \nonumber \\
&&
\left( x^{s_N} |\{s\}_y^N,-\ra - x^{-s_N}
|\{s\}_y^N,+\ra \right) = \nonumber \\
&&
\sum_{y=0}^{2^{N+1}-1} {\phi_{N+1} \over \sqrt{Z^-_{N+1}}}  e^{-\beta
H^-_{N+1}[\{s\}_y]/2} |\{s\}_y^{N+1}\ra .
\label{eq:T^-_N+1:af}
\ea

\noindent This proves that $T^-_N$ simulates the 1D antiferromagnetic
Ising model.

\subsection{The ``1D Spin-Glass'' Case}
\label{glass}

We now consider the simplest case of a random-bond Ising model with open
boundary conditions: the quenched, mixed
ferro-antiferromagnetic linear chain (also known as the $\pm J$ spin
glass). The Hamiltonian in this case may be written as:

\be
H^{\bf J}_N = -\sum_{i=1}^{N-1} J_i s_i s_{i+1} ,
\label{eq:H^J_N}
\ee

\noindent where ${\bf J} = (J_1,J_2,...,J_{N-1})$ is a fixed set of
parameters, each of which can be $\pm J$ and thus determines whether
the interaction between $s_i$ and $s_{i+1}$ is ferro- ($+$) or
antiferromagnetic ($-$). There are a total of $2^{N-1}$ ${\bf
J}$'s for the length-$N$ Ising chain, each of which can be regarded as
a different realization of quenched disorder. The operator which
simulates the corresponding Ising problem is a natural generalization of
$T^\pm_N$:

\be
T^{\bf J}_N = \left[ \prod_{i=N-1}^{1} S^{J_i}_{i,i+1}
\right] R_1 ,
\label{eq:T^J_N}
\ee

\noindent where \cite{QC:comment2}:

\be
S^{J_i}_{ij} | s_i,s_j \ra = {1 \over \sqrt{c}} \left( x^{-J_i s_i} |
s_i,s_j \ra + s_j x^{J_i s_i} | s_i,-s_j \ra \right) .
\label{eq:S^J_i}
\ee

\noindent The unitarity of $S^{J_i}_{ij}$ follows from the unitarity
of $S^+_{ij}$ and $S^-_{ij}$. Let us prove that $T^{\bf J}_N$ simulates 
the appropriate Ising problem, again, by induction: for $N=2$ there are two
realizations of the quenched disorder: $J_1 = \pm J$. Accordingly,
there are two $T^{\bf J}_2$'s: $T^+_2$ and $T^-_2$. We have already
shown that these operators solve their corresponding Ising
problem. Assume by induction that $T^{\bf J}_N$ simulates the Ising
problem for $N$ spins. There are now four
possibilities in going to $N+1$, since both $J_{N-1}$ and $J_N$ can be
$\pm J$. In fact we have already dealt with the two cases $J_{N-1} = J_N$ in
proving the algorithm for the fully ferro- and antiferromagnetic
cases. But instead of considering separately the cases $J_{N-1} \neq
J_N$, it will be more convenient to proceed generally. From the
induction hypothesis we have:

\ba
\lefteqn{ T^{\bf J}_N |\{-\}_N\ra = \sum_{y=0}^{2^N-1} A^{\bf
J}_{y} |\{s\}_y^N\ra \:\:;} \nonumber \\
&&
A^{\bf J}_{y} = 
{\phi_N \over \sqrt{Z^{\bf J}_N}} x^{-H^{\bf J}_N[\{s\}_y]} .
\label{eq:T^J_N:2}
\ea

\noindent Now, $T^{\bf J}_{N+1} = S^{J_N}_{N,N+1} T^{\bf J}_N$, so
that using Eq.(\ref{eq:S^J_i}):

\ba
\lefteqn{ T^{\bf J}_{N+1}|\{-\}_N\ra |-\ra = } \nonumber \\
&&
\sum_{y=0}^{2^N-1} {\phi_N \over \sqrt{Z^{\bf J}_N}} x^{-H^{\bf J}_N[\{s\}_y]} {1 \over \sqrt{c}} \times \nonumber \\
&&
\left( x^{-J_N s_N}
|\{s\}_y^N,-\ra - x^{J_N s_N} |\{s\}_y^N,+\ra \right) = \nonumber \\
&&
\sum_{y=0}^{2^{N+1}-1} {\phi_{N+1} \over \sqrt{Z^{\bf J}_{N+1}}}  e^{-\beta
H^{\bf J}_{N+1}[\{s\}_y]/2} |\{s\}_y^{N+1}\ra .
\label{eq:T^J_N+1}
\ea

\noindent The amplitude-squared of the configuration coded by
$\{s\}_y^{N+1}$ is exactly its thermodynamic weight for a given quenched
disorder ${\bf J}$, so this completes the proof for the mixed
ferro-antiferromagnetic case. Of course, the fully ferro- and
antiferromagnetic cases are specific instances of this general
model. The implementation of the algorithm in the present case,
according to Eq.(\ref{eq:T^J_N}), would entail using (apart from the
$\pi/2$ rotation), two different operators, in an order dictated by
the sequence of ferro/antiferromagnetic bonds in the Ising model one
wishes to solve. The complexity, however, remains $O(N)$.

The generalization to continuous interactions is straightforward. In this case:

\be
H^{\bf G}_N = -\sum_{i=1}^{N-1} G_i s_i s_{i+1} \:, \:\:\:\:\: {\bf G}
\equiv (G_1,G_2,...,G_{N-1}) ,
\label{eq:H^G_N}
\ee

\noindent where now ${\bf G}$ is a set of
independent random variables, which do not have to be identically
distributed. Suppose one prepares a set ${\bf G}$. This corresponds
to choosing a certain realization of quenched disorder in the Ising
spin glass. The quantum operator which simulates for the thermodynamic 
weights in this case is 

\be
T^{\bf G}_N = \left[ \prod_{i=N-1}^{1} S^{G_i}_{i,i+1}
\right] R_1 ,
\label{eq:T^G_N}
\ee

\noindent where:

\ba
\lefteqn{ S^{G_i}_{ij} | s_i,s_j \ra = } \nonumber \\
&&
{1 \over \sqrt{c_i}} \left( x^{-G_i s_i} | s_i,s_j \ra + s_j x^{G_i s_i}
| s_i,-s_j \ra \right) \: ; \nonumber \\
&&
c_i = 2\cosh(\beta G_i) .
\label{eq:S^G_i}
\ea

\noindent In matrix form:

\be
S^{G_i} = {1 \over \sqrt{c_i}}
\left(
\begin{array}{llcl}
 x^{G_i}            & x^{-G_i} & 0                  & 0        \\
\!\!\!\!\!-x^{-G_i} & x^{G_i}  & 0                  & 0        \\
 0                  & 0        & \:\:\:\:\:x^{-G_i} & x^{G_i}  \\
 0                  & 0        & -x^{G_i}           & x^{-G_i} \\
\end{array}
\right)
\label{eq:S^J_i-mat}
\ee

\noindent The only difference from $S^{J_i}_{ij}$ of the $\pm J$ spin
glass is that each Ising-entanglement now has its own normalization
factor, which clearly has no effect on the proof of the algorithm.

As $N$ increases finite size effects diminish. By construction, our
algorithm will provide in $O(N)$ steps a ground state
$\{s\}^*$ of a ``1D spin glass'' with probability:

\be
p^*=\exp(-\beta H_N^{\bf G}[\{s\}^*])/Z_N^{\bf G} , 
\label{eq:p*}
\ee

\noindent This probability can be made arbitrarily
close to 1 by performing the construction for low enough
temperature. If a local minimum is found instead of the ground state,
the entire process should be repeated until the ground state is
obtained. What is the average number of steps required for locating
the {\em global} minimum in this manner? After the first run one has
probability $p^*$ of having found $\{s\}^*$. If not, one failed with
probability $q = 1-p^*$ and then succeeded with probability
$p^*$, etc. Clearly the resulting distribution is geometric, and thus the
average number of runs until the global minimum is found, is $\la n
\ra = {1/p^*}$ \cite{Mises}. The total number of steps is seen to be
$O(N)/p^*$. The question of the complexity of the algorithm for
locating a ground state thus boils down to the scaling of $p^*$ with
$N$. In 1D this is in fact trivial: there are exactly two degenerate
ground states, related by $s_i \rightarrow -s_i \:\forall i$, obtained
by simply following along the chain and satisfying all bonds. Their
energy is $E=-N_B\,J$ (where $N_B = N-1$ is the number of bonds) so
$p^* = \exp(\beta N_B\, J)/Z_N$. The temperature appears here as a
control parameter: let $\Delta$ be the difference in potential energy
between a ground state and the next lowest states
$\{s\}^\dagger$. Then Eq.(\ref{eq:p*}) predicts that $p^*$ will become
dominant, since $p^*/p^\dagger =
\exp[\Delta/(k_B T)]$. This indicates that the probability of obtaining
a ground state can be made arbitrarily close to 1 \cite{QC:comment3}.
However, this is only true as long as the degeneracy $g_{ms}$ of
metastable states with energy close (of the order of the average
interaction strength $\la G \ra$) to the ground state energy remains
small in some proper sense. For higher dimensional spin glasses, it is
well known that this number is $N$-dependent. Additional complications
may arise in relation to the connectivity of the 
system in higher dimensions. We will return to this issue in later
sections.  In order to be able the discuss such systems, the problem
of closing the boundary conditions must first be discussed. This is done next.

\subsection{Closing the Boundary Conditions}
\label{closing}

It should be remarked that the algorithm as described so far is in
fact classical. A classical probabilistic computer can run the
algorithm with exactly the same efficiency simply by randomly choosing
spin $s_i$ to be up or down with a probability determined by the
thermodynamic weight of the configuration of all other $i-1$ spins and
bonds. The difference is of course that the classical computer cannot
store all $2^N$ spin configurations. However, this by itself does not
enhance the computing power since only one configuration is accessible
by measurement of the quantum register. The
effective classicality of the algorithm is due to the fact that so
far we have only employed superpositions. In Sec.\ref{interference} we
will employ the purely quantum effect of interference in order to
deal with the problem of a 1D Ising chain with
closed boundary conditions. Here we introduce the operator required
for closing a 1D chain. Such a chain has as Hamiltonian in the $\pm J$
spin glass case:

\ba
\lefteqn{ H^{\cal J}_N = -\sum_{i=1}^{N} J_i s_i s_{i+1} \: ; }
\nonumber \\
&&
s_{N+1} \equiv s_1 , \:\:\:\:\: {\cal J} \equiv (J_1,J_2,...,J_N).
\label{eq:H^J_N:closed}
\ea

\noindent A reasonable approach to closing the loop on a quantum
computer might seem to be an application of the operator $S^{J_N}_{N,1}$ after
$T^{\bf J}_N$. However, this does not work since it changes the
amplitude of $| s_1 \ra$ which was already the correct thermodynamic weight. It turns out that a
different approach is needed. Instead of working on $| s_1 \ra$, one has to first introduce a work-qubit, say
$|w\ra$, on which the Ising-entanglement operation is performed:
$S^{|J_N|}_{N,w}$. This places $|w\ra$ in a superposition of up and down
states. Closing the loop is then performed by comparing the state of
$|s_1\ra$ to that of the workbit (rather than to that of $|s_N\ra$),
which acts effectively as the fictitious spin $s_{N+1}$. If $w = s_1$,
the loop is closed, since this corresponds to $s_{N+1} = s_{1}$, as
should be. However, one is then left to wonder what to do if $w =
-s_1$. Instead of discarding this possibility as improper, it turns
out to be fruitful to adopt a more general point of view. As will be
shown in Eq.(\ref{eq:calT}), the following interpretation also holds:
If $w = s_1$, the loop is closed 
ferromagnetically [${\rm sign}(J_N)=1$]; if $w=-s_1$, the loop is closed
antiferromagnetically [${\rm sign}(J_N)=-1$]. Since the sign of the
interaction is determined randomly, we can only specify the absolute
value of $J_N$, hence the notation $S^{|J_N|}_{N,w}$. The comparison
operation can be performed by an exclusive-or (XOR):

\be
X_{ij}|s_i,s_j\ra = s_i s_j |s_i,s_i s_j\ra ,
\label{eq:xor}
\ee

\noindent or, in matrix form:

\be
X = 
\left(
\begin{array}{rrrr}
 0 & -1 &  0 & 0  \\
 1 &  0 &  0 & 0  \\
 0 &  0 & -1 & 0  \\
 0 &  0 &  0 & 1  \\
\end{array}
\right)
\label{eq:xor-mat}
\ee

\noindent Thus following $S^{|J_N|}_{N,w}$ one applies $X_{1,w}$; the
combined operation defines a new, three-qubit operator \cite{QC:comment4}:

\ba
\lefteqn{ \Omega^{|J_i|}_{i,j,w}|s_i,s_j,w\ra \equiv
X_{j,w}S^{|J_i|}_{i,w}|s_i,s_j,w\ra = } \nonumber \\
&& -s_j {1 \over \sqrt{c_i}} \left( x^{|J_i| s_i}|s_i,s_j,-s_j w\ra -
\right. \nonumber \\
&&
\left. w x^{-|J_i| s_i}|s_i,s_j,s_j w\ra \right) .
\label{eq:Omega_ijw}
\ea

\noindent The algorithm for simulating the closed chain ``1D spin glass'' can
 now be written as:

\be
T^{\cal J}_N = \Omega^{|J_N|}_{N,1,w} T^{\rm J}_N ,
\label{eq:calT^J_N}
\ee

\noindent To prove that this formula indeed yields the correct
thermodynamic weights, we may employ the result for the open-chain
case, which was proved in the previous section, and include an extra
state for the workbit. In the calculation to follow, a ``$'$'' on the
$\sum$ symbols indicates a sum over all spins except $s_1$,
and:

\be
\Phi_N = (-1)^N \prod_{i=1}^N s_i .
\label{eq:Phi}
\ee

\noindent Now:

\end{multicols}
%%\hrulefill\

\ba
\lefteqn{ T^{\cal J}_N |\{-\}_N\ra |w=-1\ra = \Omega^{|J_N|}_{N,1,w}
T^{\bf J}_N |\{-\}_N\ra |-\ra = } \nonumber \\
&&
\Omega^{|J_N|}_{N,1,w} \sum_{y=0}^{2^N-1} {\phi_N \over \sqrt{Z^{\bf J}_N}}
x^{-H^{\bf J}_N[\{s\}_y]} |\{s\}_y \ra |w\!=\!-1\ra = \nonumber \\
&&
\sqrt{2 \over c_N} \sum_{y=0}^{2^N-1} { {-s_1 \phi_N} \over \sqrt{Z^{\bf J}_N}}
x^{-H^{\bf J}_N[\{s\}_y]} |\{s\}_y \ra
\left( x^{|J_N| s_N} |w\!=\!s_1\ra + x^{-|J_N| s_N} |w\!=\!-s_1\ra
\right) = \nonumber
\ea

\noindent upon collecting terms with equal $|w\ra$:

\ba
\lefteqn{ \sqrt{2 \over {c_N\, Z^{\bf J}_N}} \sum_{y}' \Phi_N \left[
\left(
x^{-H^{\bf J}_N[\{-1,s_2,...,s_N\}_y]} x^{-|J_N| s_N}
|\{-1,s_2,...,s_N\}_y \ra + \right. \right. } \nonumber \\
&&
\:\:\:\:\:\:\:\:\:\: \left. \left. x^{-H^{\bf
J}_N[\{1,s_2,...,s_N\}_y]} x^{|J_N| s_N} |\{1,s_2,...,s_N\}_y \ra
\right) |+\ra + \right. \nonumber \\
&&
\left. \left( x^{-H^{\bf J}_N[\{1,s_2,...,s_N\}_y]} x^{-|J_N| s_N}
|\{1,s_2,...,s_N\}_y \ra + \right. \right. \nonumber \\
&&
\left. \left.  \:\:\:\:\:\:\:\:\:\: x^{-H^{\bf
J}_N[\{-1,s_2,...,s_N\}_y]} x^{|J_N| s_N} |\{-1,s_2,...,s_N\}_y \ra
\right) |-\ra \right] = \nonumber \\
&&
\sqrt{2 \over {c_N\, Z^{\bf J}_N}} \sum_{y=0}^{2^N-1}
\Phi_N x^{-H^{\bf J}_N[\{s\}_y]} |\{s\}_y \ra \left(
x^{s_N s_1 |J_N|} |+\ra + x^{-s_N s_1 |J_N|} |-\ra
\right) = \nonumber \\
&&
\sqrt{2 \over {c_N\, Z^{\bf J}_N}} \sum_{{\rm sign}(J_N)}
\sum_{y=0}^{2^N-1} 
\Phi_N x^{-H^{\bf J}_N[\{s\}_y]} x^{J_N s_N s_1} |\{s\}_y \ra
|{\rm sign}(J_N) \ra = \nonumber \\
&&
\sum_{{\rm sign}(J_N)} \sum_{y=0}^{2^N-1} {\Phi_N \over \sqrt{Z^{\cal J}_N}}
e^{-\beta H^{\cal J}_N[\{s\}_y]/2} |\{s\}_y \ra |{\rm sign}(J_N) \ra .
\label{eq:calT}
\ea

%%\hrulefill\
\begin{multicols}{2}

\noindent Since the amplitude-squared of the state $|\{s\}_y \ra$ is
given by the thermodynamic weight of the corresponding 1D,
closed-chain spin glass system, we have proved that the algorithm
which includes a workbit works for closed boundary conditions.

How should one interpret the ``$|{\rm sign}(J_N) \ra$'' in the last
line of Eq.(\ref{eq:calT})? The workbit is in the excited-
or ground state according to whether ${\rm sign}(J_N)$ is positive or
negative, respectively. That is, the state of the workbit is
determined by whether the interaction between spins $s_1$ and $s_N$ is
ferro- or antiferromagnetic. However, in the simulation of Ising
models one is interested in a specific set of bonds, so it is
necessary to be able to determine the last bond-sign. This is especially
important for higher dimensional models, where every plaquette
corresponds to a closed 1D chain. For this reason we will next present a detailed analysis of the
complexity associated with generating a single plaquette. Before doing
so, we note that a useful corollary follows from the calculation
above, given that there was no special importance to the indices of
the spins between which the loop was closed:

\newtheorem{cor}[cor0]{Corollary}
\begin{cor}
Closing the bond between $s_i$ and $s_j$ using $\Omega^{|J|}_{ijw}$ always
produces a superposition with half the states having probabilities
equal to the thermodynamic weight of the Hamiltonian with a
ferromagnetic $J_{ij}$, and the other half with antiferromagnetic
$J_{ij}$.
\label{cor}
\end{cor}

\noindent Now, the simplest way of determining the last
bond sign is to measure it, following the application of $\Omega^{|J_{N}|}_{N,1,w}$. This
irreversible, nonunitary operation collapses
the superposition of $|w\ra$ while leaving the superposed state of the
spin-qubits intact. It randomly chooses between a ferro- or
antiferromagnetic bond connecting $s_1$ and $s_N$, i.e., chooses
between ${\rm sign}(J_N)=\pm 1$. Measurements are tantamount to
classical operations, so in light of the comment at the beginning of
this section, a measurement as a means of choosing the last bond sign
leaves the algorithm in the classical realm. Therefore it may not come
as a surprise that measurements will prove to be ineffective as a
means of obtaining the desired last bond sign. Instead, one may employ interference in order to ``erase'' one of the
last bonds, and leave only the desired one. In the following two sections
we will discuss both procedures in detail.

\subsection{Measurements as a Means of Choosing the Desired Bond Sign}
\label{Z}

A measurement of the workbit collapses its superposition into either
the ferro- or antiferromagnetic state. As we now show, which last bond sign
has the higher probability depends on whether the resulting closed
chain is frustrated or not. From the result of Eq.(\ref{eq:calT}) we
have for the $\pm J$ spin glass:

\ba
\lefteqn{ \Omega^{|J_{N}|}_{N,1,w} T^{\bf J}_N |\{-\}_N\ra |-\ra = }
\nonumber \\ 
&&
\sum_{{\rm sign}(J_N)=\pm 1} \sum_{y=0}^{2^N-1} {\Phi_N \over
\sqrt{Z^{\cal J}_N}}
e^{-\beta H^{\cal J}_N[\{s\}_y]/2} |\{s\}_y \ra |{\rm sign}(J_N) \ra ,
\label{eq:calTJ}
\ea

\noindent where ${\cal J} \equiv (J_1,J_2,...,J_N)$ and $|J_i| = J$.
Let us now define the ``partial partition functions''
$Z_N^{J_N}$, i.e., $Z_N^+$ for a ferromagnetic bond between $s_1$ and
$s_N$, and $Z_N^-$ for an antiferromagnetic bond. The relative weight
of the ferro- and antiferromagnetic subspaces can then be expressed
as: $r = Z_N^+ /Z_N^-$. Without loss of generality, let us assume from now
on that an antiferromagnetic last bond results in a frustrated system
(i.e., the total number of antiferromagnetic bonds is odd), and vice
versa. Then $r$ determines the relative probability of
obtaining a frustrated or unfrustrated chain as a result of the
measurement on the workbit. It is intuitively clear that the spin
configurations of a frustrated system will generally have a higher
energy than those of the corresponding unfrustrated system, and one
would thus expect to find $Z_N^+ > Z_N^-$. In 1D this
statement can be made exact, as we now show. Separating the last bond
one finds:

\be
Z_N^{J_N} = \sum_{y=0}^{2^N-1} e^{-\beta \left( H^{\rm J}_N[\{s\}_y]-J_N s_N
s_1 \right)} ,
\label{eq:Z_N^J_N}
\ee

\noindent which can be split into two terms, corresponding to $s_1 =
s_N$ and $s_1 = -s_N$:

\ba
Z_N^{J_N} &=& \left[ \sum_{{\{s\}}_N}'  e^{-\beta \left( H^{\rm J}_N
\left[\{s\}_{s_1=s_N}\right]-J_N \right)} \right.  \nonumber \\
&+& \left. \sum_{{\{s\}}_N}' e^{-\beta \left( H^{\rm J}_N
\left[\{s\}_{s_1=-s_N}\right]+J_N \right)} \right] .
\label{eq:Z_N^J_N:1}
\ea

\noindent Defining $H_N^0[\{s\}] \equiv H^{\rm J}_N
\left[\{s\}_{s_1=s_N}\right]$ and $H_N^1[\{s\}] \equiv H^{\rm J}_N
\left[\{s\}_{s_1 = -s_N}\right]$, we can write this as:

\ba
\lefteqn{ Z_N^{J_N} = \left( x^{2J_N} Z_N^0 + x^{-2J_N} Z_N^1
\right) , } \nonumber \\
&&
Z_N^{0,1} \equiv \sum_{\{s\}}' e^{-\beta H_N^{0,1}[\{s\}]} .
\label{eq:Z_N^J_N:2}
\ea

\noindent Using this:

\be
{Z_N^+ \over Z_N^-} = {{x^{2J} Z_N^0 + x^{-2J} Z_N^1} \over {x^{-2J} Z_N^0 +
x^{2J} Z_N^1}} .
\label{eq:weight}
\ee

\noindent The ``constrained'' Hamiltonians, where $s_N=\pm s_1$ can be written as:

\ba
H_N^0 &=& -\sum_{i=1}^{N-2} J_i s_i s_{i+1} - J_{N-1} s_{N-1}
s_N \nonumber \\
&=& H_{N-1}^{\bf J} - J_{N-1} s_{N-1} s_1 \nonumber \\
H_N^1 &=& H_{N-1}^{\bf J} + J_{N-1} s_{N-1} s_1 .
\label{eq:conH}
\ea

\noindent This allows one to break up the constrained partition
functions in a manner similar to Eq.(\ref{eq:Z_N^J_N:1}):

\ba
Z_N^0 &=& \sum_{\{s\}} e^{-\beta \left( H^{\rm
J}_{N-1}[\{s\}]-J_{N-1} s_{N-1} s_1 \right)} \nonumber \\
&=& \left[ \sum_{\{s\}}' e^{-\beta \left( H^{\rm J}_{N-1}
\left[ \{s\}_{s_1=s_{N-1}}\right]-J_{N-1} \right)} \right. \nonumber \\
&+& \left. \sum_{\{s\}}' e^{-\beta \left( H^{\rm J}_{N-1}
\left[\{s\}_{s_1=-s_{N-1}}\right]+J_{N-1} \right)} \right] \nonumber \\
&=& \left( x^{2J_{N-1}} Z_{N-1}^0 + x^{-2J_{N-1}} Z_{N-1}^1 \right) .
\label{eq:Z_N^0}
\ea

\noindent Similarly,

\be
Z_N^1 = \left( x^{-2J_{N-1}} Z_{N-1}^0 + x^{2J_{N-1}} Z_{N-1}^1 \right) .
\label{eq:Z_N^1}
\ee

\noindent One may continue to split up the Hamiltonians as in
Eq.(\ref{eq:conH}). The general pattern is seen to be ($0 \leq n \leq N-3$):

\ba
Z_{N-n}^0 = 2 \left( x^{2J_{N-n-1}} Z_{N-n-1}^0 + x^{-2J_{N-n-1}}
Z_{N-n-1}^1 \right) \nonumber \\
Z_{N-n}^1 = 2 \left( x^{-2J_{N-n-1}} Z_{N-n-1}^0 + x^{2J_{N-n-1}}
Z_{N-n-1}^1 \right) .
\label{eq:Z_N-n^0,1}
\ea

\noindent Together with the obvious initial condition for a pair of
spins, $Z_2^0 = 2x^{2J_1}$, $Z_2^1 = 2x^{-2J_1}$, this defines a
recursion relation which can be solved to yield for
Eq.(\ref{eq:Z_N^J_N:2}) ($N \geq 3$):

\ba
Z_N^{J_N} = {2^{N-2} \over x^{2\left(J+\sum_{i=1}^{N-1} J_i
\right)}} \sum_{k=0}^{N-1} x^{2(J+(-1)^{k+N-1}J_N)} f_N(k) .
\label{eq:Z_N^J_N:4}
\ea

\noindent where:

\be
f_N(k) = \sum_{i_1 < i_2 < ... < i_k = k}^{N-1}
x^{4\sum_{j=1}^k J_{i_j}} \:\:\:\:;\:\: f_N(0)=1 .
\label{eq:f}
\ee

\noindent The last function generates all possible different sums of the
$J_i$'s. For example, for $N=4$ we find:

\end{multicols}
%%\hrulefill\

\be
{Z_4^+ \over Z_4^-} =
{{{ [ 1 ] +
x^{4J} \left[ x^{4J_1} + x^{4J_2} + x^{4J_3} \right] +
\left[ x^{4(J_1+J_2)} + x^{4(J_1+J_3)} + x^{4(J_2+J_3)} \right] +
x^{4J} \left[ x^{4(J_1+J_2+J_3)} \right] }}
\over
{ \left[ x^{4J} \right] +
\left[x^{4 J_1} + x^{4 J_2} + x^{4 J_3} \right] +
x^{4J} \left[ x^{4(J_1+J_2)} + x^{4(J_1+J_3)} + x^{4(J_2+J_3)} \right] +
\left[ x^{4(J_1+J_2+J_3)} \right] }} .
\label{eq:Z_4^+/Z_4^-}
\ee

%%\hrulefill\
\begin{multicols}{2}

It can be checked that this agrees with the result obtained directly
from the corresponding Hamiltonians, and indeed,
Eq.(\ref{eq:Z_N^J_N:4}) can be proved to hold by induction
\cite{us:unpublished}. The important point about
Eq.(\ref{eq:Z_N^J_N:4}) is the difference between $Z_N^+$ and
$Z_N^-$. As can be seen in Eq.(\ref{eq:Z_4^+/Z_4^-}), there is an
alternating ratio of $x^{4J}$ between groups of similar terms in $Z_N^+$ and
$Z_N^-$. ``Similar terms'' here refers to terms with the same number
of $J_i$'s, enclosed in brackets in Eq.(\ref{eq:Z_4^+/Z_4^-}). Since
there is a one-to-one correspondence of this type, we may utilize the
elementary inequality \cite{QC:inequality}

\be
\min_i \left( {a_i \over b_i} \right) \leq {\sum_i^N a_i \over \sum_i^N b_i}
\leq \max_i \left( {a_i \over b_i} \right) ,
\label{eq:ineq}
\ee

\noindent to obtain that:

\ba
\min \left( x^{4J}, x^{-4J} \right) &=& x^{-4|J|} \nonumber \\
&\leq& {Z_N^+ \over Z_N^-} \leq
x^{4|J|} = \max \left( x^{4J}, x^{-4J} \right) .
\label{eq:weight:2}
\ea

\noindent The upper bound is approached as $T \!\rightarrow\! 0$. To see
this, use Eq.(\ref{eq:Z_N^J_N:4}) to express the ratio of unfrustrated
to frustrated partition functions as:

\be
{Z_N^+ \over Z_N^-} =
{ {\sum_{k=0}^{N-1} x^{2J(1-(-1)^{k+N})} f_N(k)}
\over
{\sum_{k=0}^{N-1} x^{2J(1+(-1)^{k+N})} f_N(k)} } ,
\label{eq:Z_N^+/Z_N^-}
\ee

\noindent Since $J_N \!<\! 0$ results in a frustrated system, when $N$
is odd there must be an even number $k_e^*$ (zero included) of ferro-
and even number of antiferromagnetic bonds among the first $N\!-\!1$ $J_i$'s. When
$N$ is even, there must be an odd number $k_o^*$ of ferro- and even number of
antiferromagnetic bonds among the first $N\!-\!1$ $J_i$'s. Now, as $T \!
\rightarrow\! 0$ the dominant term in both numerator and denominator of
Eq.(\ref{eq:Z_N^+/Z_N^-}) will be the one which has all the positive $J_i$'s (such a term
exists, since $f_N(k)$ generates all combinations of $J_i$'s). When
$T$ is sufficiently low, all other terms become negligible, since they
are smaller by at least $x^{4J}$. Thus to understand the low $T$
behavior of Eq.(\ref{eq:Z_N^J_N:4}) it suffices to consider that of
the dominant terms. As can be seen from the expression for $f_N(k)$,
$k$ counts how many $J_i$'s appear in every term. Thus for odd $N$ the
dominant term is generated when $k\!=\!k_e^*$ (is even), whereas for even
$N$, the dominant term results when $k\!=\!k_o^*$ (is odd). In both cases
$k\!+\!N$ is odd. But this means that $2(J\!-\!(-1)^{k+N}J_N)$ is zero in the
frustrated ($J_N\!=\!-J$) case, $4J$ in the unfrustrated case. Thus the dominant term
in the unfrustrated case is always greater by $x^{4J}$ than that of
the frustrated case. This proves that the upper bound in
Eq.(\ref{eq:weight:2}) is indeed reached as $T \!\rightarrow\! 0$. The
implication is that an algorithm which attempts to generate
frustrated isolated plaquettes 
by measurement alone,
will have to try an average of $\sim \!
x^{4J}$ times before succeeding, and this number grows exponentially
as the temperature is lowered. 

One might be tempted to try to correct a ``wrong'' plaquette instead
of ``discarding'' it. However, it turns out that any correction
procedure has probability less than 1 of succeeding.  For example,
suppose one is interested in the frustrated case, but the measurement
yielded an unfrustrated plaquette. The problem is then that $H^{\cal
J}_N$ includes the bond $J_N$ with the wrong sign. One could imagine several
strategies to ``undo'' this, which all start with a new workbit
$w'$. For example, one could employ a ``biased random walk'' procedure:
one redoes $\Omega^{|J_N|}_{N,1,w'}$, measures again, etc. The hope is
that the resulting sum of $J_N$'s with random signs will at one point
add up to produce the originally desired ferromagnetic bond. The
probability for this to happen is equal to the probability of return of some biased random
walk where the bias increases with the distance from the origin. For the unbiased random walker in 1D
and 2D this probability is 1, but the waiting time is infinite
\cite{Hughes}. For the biased random walker the probability of return
turns out to be zero. This means that a ``random walker'' correction
procedure cannot guarantee the desired result, at constant
temperature. We are not aware of any other, more successful 
procedure.  What about turning a frustrated last bond into an
unfrustrated one?  Even this has probability less than 1, as a
consideration of a three-spin system will illustrate: Suppose one
chooses $J_1 \!=\! J_2
\!=\! J$ and after measuring $w$ one finds $J_3 \!=\! -J$. One might then
hope to correct this frustrated system by redoing
$\Omega^{|J_3|}_{3,1,w'}$. A low $T$ analysis will suffice to show
that this will not correct the error. At low $T$ the dominant spin
configuration will be that with the lowest energy $E_{min}$. It is
easily checked that both $w'=\pm 1$ yield $E_{min}\!=\!-2J$ and thus
have equal probability. The case $w'=1$ corresponds to a
ferromagnetic last bond and therefore corrects the
Hamiltonian. However, in the equally probable opposite case the
Hamiltonian now includes a last bond of strength $J_3 \!=\! -2J$. If
this had been the result of the correction procedure, one would be facing a
(wrongly) biased random walk again, since a new measurement would
either result in $J_3
\!=\! J$ or $J_3 \!=\! -3J$, with corresponding $E_{min}\!=\!-J$ and
$E_{min}\!=\!-3J$. The latter is more probable by a factor $e^{2\beta
J}$, so the correction fails.

These arguments show that procedures using only
combinations of superposition and measurement, have no control over
the type of plaquette they generate. In the next section we will
introduce a procedure which does have this feature.

\subsection{Using Interference to Close an Isolated Plaquette}
\label{interference}

After closing the last bond, the quantum register is in a
superposition corresponding to a frustrated ($w=-1$) and unfrustrated ($w=1$)
plaquette [Eq.(\ref{eq:calT})] (assuming, without loss of generality,
that there is an even number of antiferromagnetic bonds among the
first $N-1$ bonds):

\be
|\psi\ra = T^{\cal J}_N |\{-\}_N\ra |w=-1\ra =
|\psi_-\ra + |\psi_+ \ra ,
\label{eq:psi}
\ee

\noindent where

\ba
|\psi_{\pm} \ra = {1 \over \sqrt{Z^{\cal J}_N}} 
\sum_{y=0}^{2^N-1} \Phi_N x^{-H_{\pm}[\{s\}_y]}
|\pm \ra |\{s\}_y\ra ,
\label{eq:psi+-}
\ea

\noindent and where $H_{\pm}$ corresponds to $J_N=\pm J$. We have
intentionally written the workbit first, so that in the binary
representation of the spin configurations by the quantum register, the
first $2^N$ states correspond to the frustrated configurations
($w\!=\!-1$) and the last $2^N$ correspond to the unfrustrated configurations
($w\!=\!+1$). As was shown
in the previous section, achieving control 
over the plaquette type cannot be done by measurement alone. However,
one may try to employ interference in order to ``erase'' one of the
subspaces, thus leaving only the desired plaquette type. To see this,
consider the wavefunction $|\psi\ra$ of the quantum register as a
vector of length $2^{(N+1)}$, with the first $2^N$ entries 
corresponding to $|\psi_-\ra$, and the last $2^N$ entries 
corresponding to $|\psi_+\ra$. Within each such subspace, the entries run
over all possible spin configurations $y=0...2^N-1$. Clearly $\la
\psi | \psi \ra \!=\! 1$ by unitarity of $T^{\cal J}_N$. We now seek a
new unitary transformation $U$ such that:

\be
U_{\pm} |\psi\ra = {1 \over \sqrt{Z_{\pm}}} |\psi_{\pm}\ra
\:\:\:\:\:;\:\:
Z_{\pm} = \la \psi_{\pm} |\psi_{\pm} \ra .
\label{eq:U}
\ee

\noindent Thus $U$ rotates the superposed quantum register state
$|\psi\ra$ into a state representing either the frustrated or the
unfrustrated configurations. That $U$ exists is clear since it takes
one norm 1 vector into another. Furthermore it clearly mixes different
spin configurations, i.e., creates interferences. The problem is to
find $U$, given that it is a $2^{(N+1)}\cdot 2^{(N+1)}$ matrix of
coefficients that depend on $\cal J$. In other words, one needs to
know the Gibbs distribution of the plaquette as input in order to find
$U$! This might seem to defeat the purpose of the algorithm
altogether, but not so in view of the need to construct plaquettes
with given bonds for the 2D and 3D Ising problems. In these higher
dimensional cases it is very useful to know how to produce small
plaquettes of, e.g., 3 or 4 spins, for the triangular and square
lattices respectively. Thus in these cases, or indeed for any
plaquette of $N$ spins, one could calculate in advance the Gibbs
distribution, find $U$, and use it in the construction of a
lattice. We will deal with the question of ``integration'' of a
plaquette into a lattice in Sec.\ref{2D}. Here we give the general
(classical) algorithm for the construction of $U$ for any $N$, and
explicitly solve the cases $N=3,4$.\\

\subsubsection{General Construction of $N$-Spin Interference Operator}
Let $\{\hat{e}_{i}\}_{i=1}^{2n}$, $n=2^N$, be the standard basis of vectors for
${\cal R}^{2n}$, with a 1 at position $i$, zeroes elsewhere. Consider
a normalized real vector ${\bf v}$ of length $2n$ (representing
$|\psi\ra$) and another vector ${\bf w}$ composed of ${\bf v}$'s upper
half, also normalized: ${\bf w} = (v_1,v_2,...,v_n,0,0,...,0)/\alpha_-$ where
$\alpha_- = \left(\sum_{i=1}^n v_i^2\right)^{1/2}$. Here ${\bf w}$ corresponds to
$|\psi_-\ra$. We seek a construction by two-qubit
operations of a unitary matrix $U_{\_}$ such that $U_{\_}{\bf v} = {\bf w}$ [as
in Eq.(\ref{eq:U})]. The solution to this problem is within the theory
of generators of $SO(n)$, as outlined in Ref.\cite{Vilenkin}, in terms
of generalized Euler angles. Since in our case ${\bf v}$ and ${\bf w}$
are real vectors, it suffices to find an orthogonal $U_{\_}$. The idea is
to first rotate ${\bf w}$ so that it coincides with $\hat{e}_{n}$, and
then solve the easier problem of finding the transformed $U_{\_}$ which
rotates the transformed ${\bf v}$ into $\hat{e}_{n}$. More explicitly,
suppose one has found an orthogonal matrix $G_{-}^{(1)}$ satisfying: ${\bf w}
\!=\! G_{-}^{(1)} \hat{e}_{n}$. Clearly $G_{-}^{(1)}$ is composed of two blocks, an
upper block $G_-'^{(1)}$ and a lower one $I_n$, the $n\!\cdot\! n$ identity
matrix. The transformed equation is then: $G_{-}^{(2)} {\bf v}' = \hat{e}_{n}$, with $G_{-}^{(2)} \!=\! (G_{-}^{(1)})^{-1} U_{\_} G_{-}^{(1)}$ and:

\ba
{\bf v}' &=& (G_{-}^{(1)})^{-1}{\bf v} =
(G_{-}^{(1)})^{-1}(\alpha_-{\bf w} +
\sum_{i=n+1}^{2n} v_i\hat{e}_{i}) \nonumber \\
&=& (\overbrace{0,0,..,\alpha_-}^{n},v_{n+1},...,v_{2n}) .
\label{eq:v'}
\ea

\noindent Since only the last $n\!+\!1$ coordinates of ${\bf v}'$ are non-zero, $G_{-}^{(2)}$ is
composed of an upper block $I_{n-1}$ and a lower block $G_-'^{(2)}$ which we
need to find along with $G_-'^{(1)}$. Having found these, the solution can
be written as:

\be
U_{\_} \!=\! G_{-}^{(1)} G_{-}^{(2)} (G_{-}^{(1)})^{-1} .
\label{eq:U-G}
\ee

\noindent Following Ref.\cite{Vilenkin}, let us write:

\be
G_{-}^{(1)} = \prod_{i=1}^{n-1} g_i(\theta_i), \:\:\:\:
G_{-}^{(2)} = \prod_{i=n}^{2n-1} g_i(\theta_{i}),
\label{eq:G-12}
\ee

\noindent where
$g_i(\theta_i)$, is a rotation by $\theta_i$ in the
plane spanned by $(\hat{e}_{i},\hat{e}_{i+1})$ of ${\cal R}^{2n}$:

\be
g_i(\theta_i) =
\left(
\begin{array}{crcc}
 I_{i-1} &               &              &            \\
         & \cos\theta_i  & \sin\theta_i &            \\
         & -\sin\theta_i & \cos\theta_i &            \\
         &               &              & I_{2n-i-1} \\
\end{array}
\right) ,
\label{eq:g_i}
\ee

\noindent Application of $G_{-}^{(1)}$ to $\hat{e}_{n}$ results in the
following set of equations: 

\ba
\lefteqn{w_n = \cos\theta_{n-1}} \nonumber \\
&&
w_{n-1} = \sin\theta_{n-1} \cos\theta_{n-2} \nonumber \\
&&
w_{n-2} = \sin\theta_{n-1} \sin\theta_{n-2} \cos\theta_{n-3} \nonumber \\
&&
\cdots \nonumber \\
&&
w_{2} = \sin\theta_{n-1} \sin\theta_{n-2}
\cdots \sin\theta_{2} \cos\theta_{1} \nonumber \\
&&
w_{1} = \sin\theta_{n-1} \sin\theta_{n-2} \cdots
\sin\theta_{2} \sin\theta_{1} ,
\label{eq:w}
\ea

\noindent with solution ($k=1,...,n-1$):

\ba
\lefteqn{
\cos\theta_k = {v_{k+1} \over {r_{k+1}} } } \nonumber \\
&&
\sin\theta_k = {r_{k} \over {r_{k+1}} } \nonumber \\
&&
r_{j} = \left(\sum_{i=1}^{j} v_{i}^2\right)^{1/2}
\label{eq:wcossin}
\ea

\noindent Similarly, application of $(G_-^{(2)})^{-1}$ to
$\hat{e}_{n}$ results in:

\ba
\lefteqn{v'_n = \cos\theta_n} \nonumber \\
&&
v'_{n+1} = \sin\theta_{n} \cos\theta_{n+1} \nonumber \\
&&
v'_{n+2} = \sin\theta_{n} \sin\theta_{n+1}
\cos\theta_{n+2} \nonumber \\
&&
\cdots \nonumber \\
&&
v'_{2n-1} = \sin\theta_{n} \sin\theta_{n+1}
\cdots \sin\theta_{2n-2} \cos\theta_{2n-1} \nonumber \\
&&
v'_{2n} = \sin\theta_{n} \sin\theta_{n+1} \cdots
\sin\theta_{2n-2} \sin\theta_{2n-1} ,
\label{eq:v':1}
\ea

\noindent yielding ($k=n,...,2n-1$):

\ba
\lefteqn{
\cos\theta_k = {v'_{k} \over r_{k}} } \nonumber \\
&&
\sin\theta_k = {r_{k+1} \over r_{k}}\:, \:\:\: \sin\theta_{2n-1} =
-{r_{k+1} \over r_{k}} \nonumber \\
&&
r_{j} = \left(\sum_{i=j}^{2n} v_{i}'^2\right)^{1/2} .
\label{eq:vcossin}
\ea

\noindent The case ${\bf w} \!=\!
(0,...,0,v_{n+1},...,v_{2n})/\alpha_+$, with $\alpha_+ = \left(\sum_{i=1}^n
v_i^2 \right)^{1/2}$, corresponds to $|\psi_+\ra$, and we need to find
an orthogonal $U_+$ such that $U_+ {\bf v} \!=\! {\bf w}$. It
can be seen by repeating the arguments above that we now require ${\bf
w} \!=\! G_{+}^{(1)} \hat{e}_{n+1}$, whence $G_{+}^{(1)}$ has $I_n$
as its upper block, and ${\bf v}' \!=\! G_{+}^{(1)} {\bf v} \!=\!
(v_1,...v_n,\alpha_+,0,...,0)$ satisfies the transformed equation
${\bf v}' \!=\! G_{+}^{(2)} \hat{e}_{n+1}$, where $G_{+}^{(2)} \!=\!
G_{+}^{(1)} U_+^{-1} (G_{+}^{(1)})^{-1}$ has $I_{n-1}$ as its lower
block. Writing accordingly

\be
G_{+}^{(1)} = \prod_{i=n+1}^{2n-1} g_i(\theta_i), \:\:\:\:
G_{+}^{(2)} = \prod_{i=1}^{n} g_i(\theta_{i}),
\label{eq:G+12}
\ee

\noindent leads to equations very similar to
Eqs.(\ref{eq:w}-\ref{eq:vcossin}): one needs to replace $w_i$ and
$v_i$ by
$v'_i$ everywhere in Eqs.(\ref{eq:w}-\ref{eq:wcossin}), as well as
allow $k$ to range from 1 to $n$. In Eq.(\ref{eq:v':1}) one should
replace $v'_i$ by $w_i$, whereas in Eq.(\ref{eq:vcossin}) $k$ should range from
$n+1$ to $2n-1$ and $v'_i$ replaced by $v_i$. With these replacements
it follows that the $\theta_i$ are identical for $U_-$ and $U_+$,
except for $\theta_n$, for which $\sin\theta_n^+ \!=\!
\cos\theta_n^-$. The interference matrix is given by:

\be
U_{+} \!=\! (G_{+}^{(1)})^{-1} G_{+}^{(2)} G_{+}^{(1)} .
\label{eq:U+G}
\ee

\noindent This, together with Eqs.(\ref{eq:v'}),(\ref{eq:U-G}),
uniquely solves our problem. It remains to be shown explicitly how
$U_{\pm}$ can be written in terms of one and two-qubit operations on
the original spin-qubit register.

We note that the $g_i(\theta_i)$ are identity matrices
except for $2\!\cdot\! 2$ blocks. Consider the representation of the basis
$\{\hat{e}_i\}$ by
${\bf s}' \!=\! \{|s'_1,...,s'_N\ra\}_{y'=0}^{2^N-1}$ where the successive register
states differ by a {\em single} qubit flip. In this basis application of each
$g_i(\theta_i)$ does not entangle states differing by more than
one qubit, so the corresponding $\hat{g}_i(\theta_i)$ is a single-qubit operator:

\ba
\lefteqn{
\hat{g}_i(\theta_i) |s_i\!=\!-1\ra =  \cos\theta_i |-\ra + \sin\theta_i |+\ra
} \nonumber \\
&&
\hat{g}_i(\theta_i) |s_i\!=\!+1\ra = -\sin\theta_i |-\ra + \cos\theta_i |+\ra .
\label{eq:ghat}
\ea

\noindent The remaining problem is thus seen to be the transformation from the
original ``binary'' basis $\{|s_1,...,s_N\ra\}_{y=0}^{2^N-1}$ to
${\bf s}'$ and back. This is easily accomplished by two-qubit operations using
the well-known classical Gray code \cite{Kostopoulos}. The
Gray-to-binary transformation is accomplished by successive XORs
starting from  the last qubit [these differ somewhat from the
definition in Eq.(\ref{eq:xor}), hence the different notation]:

\be
X'_{i,j} |s_i,s_j\ra \rightarrow |s'_i,s'_j\ra = |s_i, -s_i s_j\ra ,
\label{eq:gray}
\ee

\noindent where an extra workbit $w_x$ represent $s_{0} \equiv -1$. For
example, for the two successive binary ${\bf s}$ basis states
$|w_x,s_1,s_2,s_3,s_4\ra \!=\!
|-,-,+,+,+\ra$ and $|-,+,-,-,-\ra$
we find after application of $\prod_{i=0}^3 X'_{i,i+1}$: $
|-,-,+,-,-\ra$ and $|-,+,+,-,-\ra$
respectively, which indeed differ by only a single
qubit. Furthermore, clearly $X'^{-1}_{i,j} \!=\! X'_{i,j}$ so the
binary-to-Gray transformation is accomplished by running the same
sequence of XORs in reverse order (starting from the extra
workbit). We can now finally write down the full interference
transformation. Let:

\be
X'_N \equiv X'_{w_x,w} X'_{w,1} \prod_{i=1}^{N-1} X'_{i,i+1} ,
\label{eq:X'}
\ee

\noindent and:

\ba
\lefteqn{
\hat{G}_-^{(1)} = \prod_{i=1}^{2^N-1} \hat{g}_i(\theta_i) \:\:\:\:\:\:\:
\hat{G}_-^{(2)} = \prod_{i=2^N}^{2^{N+1}-1} \hat{g}_i(\theta_{i})
} \nonumber \\
&&
\hat{G}_+^{(1)} = \prod_{i=2^N+1}^{2^{N+1}-1} \hat{g}_i(\theta_{i}) \:\:\:
\hat{G}_+^{(2)} = \prod_{i=1}^{2^N} \hat{g}_i(\theta_{i}) .
\label{eq:Ghat}
\ea

\noindent The inverse operators are obtained by reversing the order of
the products and negating all angles. Then:

\ba
\lefteqn{
|w_x\ra |\psi_-\ra = X'^{-1}_N {\hat{G}_-}^{(1)} {\hat{G}_-}^{(2)}
({\hat{G}_-}^{(1)})^{-1} X'_N |w_x\ra |\psi\ra
} \nonumber \\
&&
|w_x\ra |\psi_+\ra = X'^{-1}_N ({\hat{G}_+}^{(1)})^{-1}
({\hat{G}_+}^{(2)})^{-1} {\hat{G}_+}^{(1)} X'_N |w_x\ra |\psi\ra
\label{eq:psi-}
\ea

\noindent expresses the interference transformation leaving only the
register states corresponding to frustrated or unfrustrated spin
configurations. \\

\subsubsection{Solution for $N=3,4$ Spin System}
We now employ the above formalism in order to explicitly solve for the
interference transformations of the three- and four-spin systems,
corresponding to the elementary cells of the triangular and square lattices respectively. First it should
be noted that for the $\pm J$ model in 1D, for a closed Ising chain of given
length, the spectra of all frustrated realizations of bond choices are
identical, and so are the spectra of all unfrustrated realizations. To
see this, consider a specific spin configuration and realization of
bonds with energy $E$, and suppose one changes the sign of some arbitrary {\em pair} of
bonds $J_m, J_n$, $m<n$. This operation does not change the
frustration of the chain, since this is determined by the parity of
antiferromagnetic bonds. But by flipping all spins $s_{m+1},...,s_n$
once again the energy is $E$, since the flipping of $s_{m+1}$ and
$s_n$ undoes the change in sign of $J_m$ and $J_n$ respectively, and
all other spin flips occur in pairs that share a bond and thus cancel. So for every
spin configuration and choice of bonds, there is another spin
configuration with the same energy in a realization with the same
frustration but different bonds. Clearly the mapping above is
one-to-one, so that indeed the spectrum is identical for all bond
realizations with the same frustration. Returning to the $N=3,4$ spin
systems, we are at liberty to consider, e.g., the case where all bonds
but the last are ferromagnetic. The superposition of $J_N$ into a ferro- and
antiferromagnetic bond will represent all other unfrustrated and
frustrated bond realizations, respectively. Solution of
Eqs.(\ref{eq:wcossin}),(\ref{eq:vcossin}) yields the following result
for the transformation from the superposition to the frustrated or unfrustrated
subspaces. Let:

\ba
\lefteqn{
f_{k,l,m} \equiv {x^{4J k} \over \sqrt{l + m x^{8J}}} } \nonumber \\
&&
h_{k,l,m,n} \equiv {x^{4J k} \over \sqrt{l + m x^{8J} + n x^{16J}}} .
\label{eq:fh}
\ea

\noindent Then, writing $C\equiv\cos$, the angles can be expressed as:

\end{multicols}

\begin{table}
\begin{tabular}{lllllll}
$N=3$: & & & & & & \\
$C\theta_1 = {1 \over \sqrt{2}}$ &
$C\theta_2 = -{1 \over \sqrt{3}}$ &
$C\theta_3 = -f_{0,1,3}$ &
$C\theta_4 = f_{1,1,4}$ &
$C\theta_5 = f_{1,1,5}$ &
$C\theta_6 = -f_{1,1,6}$ &
$C\theta_7 = -f_{0,2,6}$ \\
$C\theta_{15} = -{1 \over \sqrt{2}}$ &
$C\theta_{14} = {1 \over \sqrt{3}}$ &
$C\theta_{13} = f_{1,1,3}$ &
$C\theta_{12} = -f_{0,4,1}$ &
$C\theta_{11} = -f_{0,5,1}$ &
$C\theta_{10} = f_{0,6,1}$ &
$C\theta_9 = f_{1,6,2}$ \\
$C\theta_8^- =$ & $\left(f_{0,1,3} (1+x^{4J})^{3/2}\right)^{-1}$ &
$= \sin\theta_8^+$ & & & & \\ \hline
$N=4$: & & & & & & \\
$C\theta_1 = {1 \over \sqrt{2}}$ &
$C\theta_2 = -{1 \over \sqrt{3}}$ & & & & & \\
$C\theta_3 = -h_{0,1,3,0}$ &
$C\theta_4 = h_{1,1,4,0}$ & & & & & \\
$C\theta_5 = h_{0,2,4,0}$ &
$C\theta_6 = -h_{0,3,4,0}$ &
$C\theta_7 = -h_{0,4,4,0}$ & & & & \\
$C\theta_8 = -h_{1,4,5,0}$ &
$C\theta_9 = -h_{1,4,6,0}$ &
$C\theta_{10} = h_{1,4,7,0}$ & & & & \\
$C\theta_{11} = h_{0,5,7,0}$ &
$C\theta_{12} = -h_{1,5,8,0}$ & & & & & \\
$C\theta_{13} = -h_{0,6,8,0}$ &
$C\theta_{14} = h_{0,7,8,0}$ &
$C\theta_{15} = h_{0,8,8,0}$ & & & & \\
$\sin\theta_{16}^- =$ & $\left(h_{0,1,6,1} (1+x^{4J})^2\right)^{-1}$ &
$= C\theta_{16}^+$ & & & & \\
$C\theta_{17} = -h_{2,2,12,2}$ &
$C\theta_{18} = -h_{1,2,12,1}$ & & & & & \\
$C\theta_{19} = h_{1,2,11,1}$ &
$C\theta_{20} = h_{1,2,10,1}$ &
$C\theta_{21} = -h_{1,2,9,1}$ &
$C\theta_{22} = -h_{1,2,8,1}$ &
$C\theta_{23} = h_{1,2,7,1}$ &
$C\theta_{24} = h_{0,2,6,1}$ \\
$C\theta_{25} = h_{2,1,6,1}$ & & & & & & \\
$C\theta_{26} = h_{1,1,6,0}$ &
$C\theta_{27} = -h_{1,1,5,0}$ &
$C\theta_{28} = -h_{1,1,4,0}$ &
$C\theta_{29} = h_{1,1,3,0}$ &
$C\theta_{30} = h_{1,1,2,0}$ &
$C\theta_{31} = -h_{1,1,1,0}$ 
\end{tabular}
\label{eq:cF4}
\end{table}

%%\hrulefill\

\begin{multicols}{2}

\nopagebreak
\noindent The regularity and similarity between terms in the same
column ($\cos\theta_i$,
$\cos\theta_j$ with $i\!+\!j \!=\! 16$) for $N=3$ is noteworthy. For the
four-spin system there is a similarity between terms in the same row,
but we find a less regular solution. 
Let us remind the reader at this point the motivation for
introducing the above transformations. We showed in Sec.\ref{Z} that
the average number of attempts needed to generate a plaquette of given
type using only 
superpositions and no interference, grows as $x^{4J} \!=\!
\exp(2J/k_B T)$ with the
temperature. Using the interference transformations, the cost is
$O(1)$ in the plaquette size, and independent of the temperature. 
We are now
finally ready to discuss more interesting Ising models, in 2D and above.

\section{Higher Dimensional Ising Models}
\label{highD}

The ``1D Ising spin glass'' is rather trivial, and the more interesting
models are the higher-dimensional ones, where connectivity plays an
important role. As an introduction to the schemes we will need to
employ in dealing with the 2D and 3D cases, consider first the
``infinite-dimensional'' Bethe lattice, which has no closed loops.

\subsection{The Bethe Lattice Case}
\label{Bethe}

\begin{figure}
\hspace{2em}
\epsfysize=8cm
\epsfxsize=8cm
\epsffile{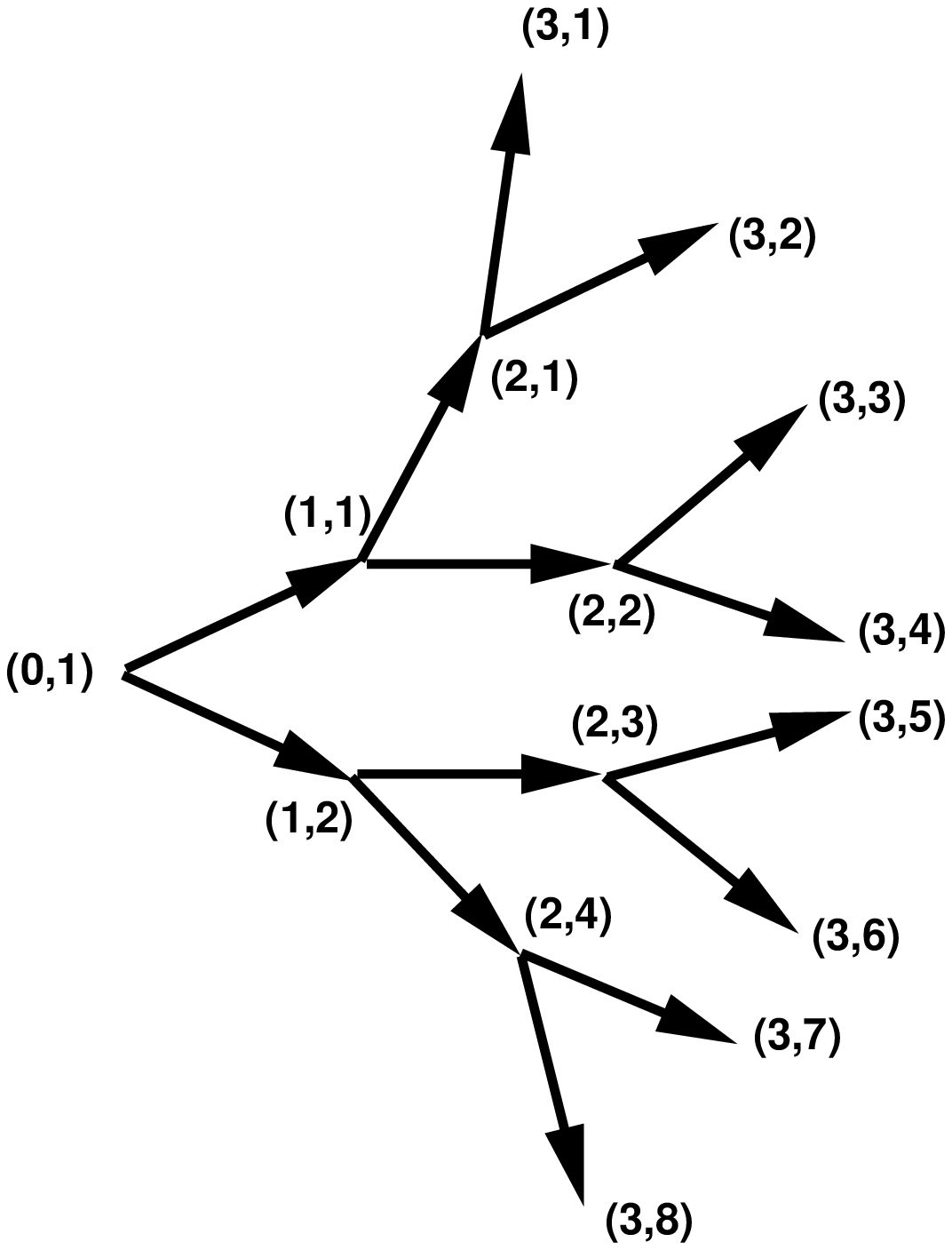}
\caption{Scheme for numbering vertices on the binary Bethe tree, used
in the Hamiltonian of Eq.(\protect\ref{eq:H-Bethe}).}
\label{fig:Bethe}
\end{figure}

Consider a binary Bethe lattice, i.e., with spins located at the
vertices of a binary tree (Fig.\ref{fig:Bethe}). The spin glass
Hamiltonian for a $K$-level deep tree can be written as:

\be
H^{\bf G}_K = -\sum_{k=1}^K \sum_{i=1}^{2^k} G_{(k-1,\lceil i/2
\rceil)(k,i)}s_{(k-1,\lceil i/2 \rceil)} s_{(k,i)} .
\label{eq:H-Bethe}
\ee

\noindent According to the general recipe of Sec.\ref{ferro}, the
quantum operator for calculating the weights of configurations in this
system is (for simplicity of notation we shall suppress the indices on
$G$ where they are already indicated by $S$):

\be
T^{\bf G}_K = \left[ \prod_{k=K}^1 \prod_{i=1}^{2^k} S^G_{(k-1,\lceil
i/2 \rceil),(k,i)} \right] R_{(0,1)} .
\label{eq:T-Bethe}
\ee

\noindent In order to prove that $T_K$ correctly calculates the
probabilities of the spin glass Ising model on the Bethe tree, we need
to show (I) that it does not matter in which order we connect the
spins occupying vertices one level deeper then their common
originator, and (II) that a 1D chain which splits at its end into two
branches is correctly described. For, (I) allows us to perform the
first branching in the tree [from spin $(0,1)$], and (II) [combined
with (I)] allows us to build up the tree recursively from any existing
end point. In particular, the order described in Eq.(\ref{eq:T-Bethe})
will be valid. Starting with (I), we need to show that
$[S^G_{ij},S^G_{ik}]=0$ (the indices $i,j,k$ are shorthand for the
double indices employed above). Now,

\end{multicols}
%%\hrulefill\

\ba
\lefteqn{ S^G_{ij}S^G_{ik}|s_i,s_j,s_k\ra = } \nonumber \\
&&
{1 \over \sqrt{c_{ik}}} S_{ij}\left[ x^{-G_{ik}s_i}|s_i,s_j,s_k\ra + s_k
x^{G_{ik}s_i} |s_i,s_j,-s_k\ra \right] = \nonumber \\
&&
{1 \over {\sqrt{c_{ij}c_{ik}}}} \left\{ x^{-G_{ik}s_i}\left[
x^{-G_{ij}s_i}|s_i,s_j,s_k\ra + s_j x^{G_{ij}s_i} |s_i,-s_j,s_k\ra
\right] + \right. \nonumber \\
&&
\left. s_k x^{G_{ik}s_i}\left[ x^{-G_{ij}s_i}|s_i,s_j,-s_k\ra + s_j
x^{G_{ij}s_i} |s_i,-s_j,-s_k\ra \right] \right\} = \nonumber \\
&&
{1 \over {\sqrt{c_{ij}c_{ik}}}} \left[ x^{-(G_{ik}+G_{ij})s_i}
|s_i,s_j,s_k\ra + s_j x^{(G_{ij}-G_{ik})s_i} |s_i,-s_j,s_k\ra +
\right. \nonumber \\
&&
\left. s_k x^{(G_{ik}-G_{ij})s_i} |s_i,s_j,-s_k\ra + s_k s_j
x^{(G_{ik}+G_{ij})s_i} |s_i,-s_j,-s_k\ra \right] .
\label{eq:SijSik}
\ea

\noindent But on the other hand, by exchanging $j$ and $k$ everywhere
in the last line, we obtain:

\ba
\lefteqn{ S^G_{ik}S^G_{ij}|s_i,s_j,s_k\ra = } \nonumber \\
&&
{1 \over {\sqrt{c_{ik}c_{ij}}}} \left[ x^{-(G_{ij}+G_{ik})s_i}
|s_i,s_k,s_j\ra + s_k x^{(G_{ik}-G_{ij})s_i} |s_i,-s_k,s_j\ra +
\right. \nonumber \\
&&
\left. s_j x^{(G_{ij}-G_{ik})s_i} |s_i,s_k,-s_j\ra + 
s_j s_k x^{(G_{ij}+G_{ik})s_i} |s_i,-s_k,-s_j\ra \right] .
\label{eq:SikSij}
\ea

%%\hrulefill\
\begin{multicols}{2}

\noindent The order of the qubits in the kets is immaterial, so that
by comparing the two results we find that indeed

\be
[S^G_{ij},S^G_{ik}]=0 .
\label{eq:[Sij,Sik]}
\ee

\noindent This is indicated graphically in
Fig.\ref{fig:rules}(a). Next we prove (II) above, namely that
$S^G_{N,N+2}S^G_{N,N+1} T^{\bf G}_N |\{-\}_N\ra
|-\ra_{N+1}|-\ra_{N+2}$ (where due to Eq.(\ref{eq:[Sij,Sik]}) we may
exchange the order of $S^G_{N,N+2}$ and $S^G_{N,N+1}$) yields the
correct thermodynamic weight for the Hamiltonian

\ba
H &=& \sum_{i=1}^{N-1} G_{i,i+1} s_i s_{i+1} + G_{N,N+1} s_N s_{N+1}
\nonumber \\
&+& G_{N,N+2} s_N s_{N+2} .
\ea

\noindent Using the results of
Eqs.(\ref{eq:T^J_N:2}),(\ref{eq:SijSik}) we have:

\end{multicols}
%%\hrulefill\

\ba
\lefteqn{ S^G_{N,N+2}S^G_{N,N+1} T^{\bf G}_N |\{-\}_N\ra
|-\ra_{N+1}|-\ra_{N+2} = } \nonumber \\
&&
\sum_{y=0}^{2^N-1} A^{\bf G}_{\yb_N} S^G_{N,N+2}S^G_{N,N+1}
|\yb_N\ra |-,-\ra =  \nonumber \\
&&
{1 \over {\sqrt{c_{N,N+2}c_{N,N+1}}}} \sum_{y=0}^{2^N-1} A^{\bf
G}_{\yb_N} |\yb_N\ra \left(
x^{-s_N(G_{N,N+1}+G_{N,N+2})} |-,-\ra - 
x^{s_N(G_{N,N+1}-G_{N,N+2})} |+,-\ra \right. \nonumber \\
&&
\left. - x^{s_N(G_{N,N+2}-G_{N,N+1})} |-,+\ra +
x^{s_N(G_{N,N+1}+G_{N,N+2})} |+,+\ra \right) =  \nonumber \\
&&
{1 \over {\sqrt{c_{N,N+2}c_{N,N+1}}}} \sum_{y=0}^{2^{N+1}-1} A^{\bf
G}_{\yb_N} s_{N+1} s_{N+2} x^{G_{N,N+1}s_N s_{N+1} + G_{N,N+2}s_N
s_{N+2}} |\yb_N\ra |s_{N+1}, s_{N+2}\ra =  \nonumber \\
&&
{1 \over {\sqrt{c_{N,N+2}c_{N,N+1}}}} \sum_{y=0}^{2^{N+1}-1} {1 \over
\sqrt{Z^{\bf G}_N}} e^{-\beta \left( H^{\bf G}_N [\{s\}_y] - G_{N+1} s_N
s_{N+1} - G_{N+2} s_N s_{N+2} \right) } |\yb_{N+2}\ra ,
\label{eq:Bethe-calc}
\ea

%%\hrulefill\
\begin{multicols}{2}

\noindent which is the desired result. It should be noted that since
$[S^G_{ij},S^G_{ik}]=0$, any number of $S$ operators with a common
starting point $i$ will commute in pairs. Therefore there is nothing
special about the binary Bethe tree, and we can equally well apply our
algorithm, after a proper modification of Eq.(\ref{eq:T-Bethe}), to
a Bethe tree with any kind of branching.

\subsection{2D Ising Model}
\label{2D}

As was demonstrated in the 1D case, the key to being able to close
loops is the creation of a superposition in {\em bond}-space by using
a workbit whose state is compared with the spin with which the loop is
closed. Choosing a specific bond-sign is then 
accomplished by an interference transformation which eliminates one of
the bond subspaces. We now extend these ideas in order to present an
algorithm for simulating 2D Ising spin systems. Ideally one would like
to have an algorithm which can exactly calculate the thermodynamic
weights of an arbitrary given spin glass Hamiltonian:

\be
H = - \sum_{\la i,j \ra} J_{ij}s_i s_j \:\: ;\:\:\:\:\: |J_{ij}| = J .
\label{eq:H-SG}
\ee

\noindent However, since the interference transformations introduced
in the 1D case require as input the thermodynamic weights, we cannot
hope to deal with an arbitrary Hamiltonian. Instead, as will be shown
here, the class of spin glasses which can be dealt with by the present
algorithm is that with predetermined plaquettes of finite size. In
other words, by using interfence transformations one can construct
{\em isolated} plaquettes of any desired (finite) size and composition
(of bonds), and these plaquettes can then be connected together. This
creates new plaquettes, with random bond signs. 
Thus the algorithm cannot
provide complete control over the bond composition of the Ising model
it is used to simulate, but the resulting class of partially
random-bond systems is huge (exponentially large in the number of
bonds).
Furthermore, by
lowering the temperature one can increase the probability of
generating only unfrustrated plaquettes connecting the prefabricated
ones. For example, choosing 
prefabricated unfrustrated plaquettes will generate low-temperature
simulations of unfrustrated Ising models with very few defects. That
is, if $N_d$ denotes the number of defect (i.e., frustrated) plaquettes, then:

\be
{N_d \over N} \sim x^{-4J} .
\ee

\noindent We turn next to demonstrating how isolated plaquettes can be
connected 
together to form a lattice.

\subsubsection{Allowed Algorithms}
\begin{figure}
\hspace{2em}
\epsfysize=9cm
\epsfxsize=9cm
\epsffile{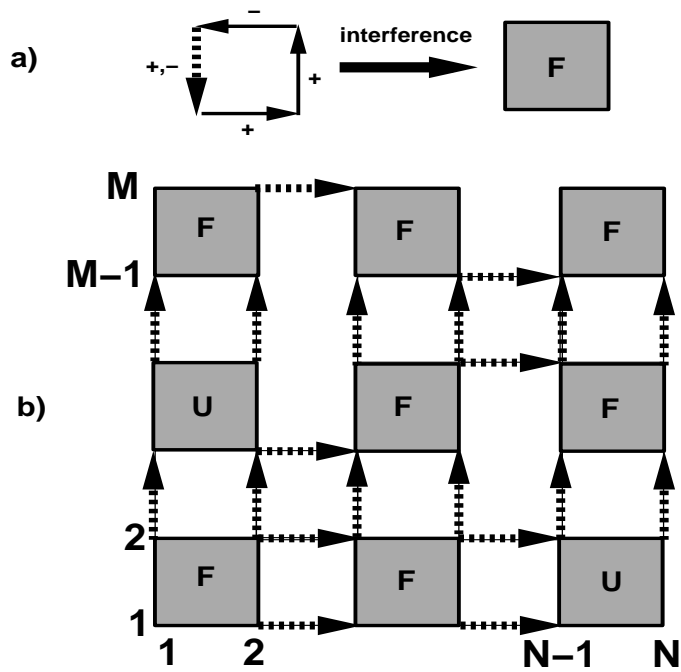}
\caption{a) Preparation of a single plaquette by three $S$ operators
(ferro (+), ferro, antiferro (-)) and closing with $\Omega$. This is
followed by an interference 
transformation which erases the antiferromagnetic subspace, leaving in
this case a frustrated (F) plaquette. b) A lattice with full density
of prefabricated plaquettes (some unfrustrated -- U), connected by $\Omega$ operators (dashed
arrows), before measurement of the bonds. In this case all vertical bonds are present and the lattice
is diluted in the horizontal bonds. }
\label{fig:2D}
\end{figure}

``Hooking up'' isolated plaquettes will require connecting spins by
$\Omega$ operators, all of which will eventually have to share lattice
points in pairs (or more), as shown in Fig.\ref{fig:2D}. Corollary \ref{cor} assures that bonds can
always be closed using $\Omega$ so as to produce the correct superposition. Since the order in which the lattice is
constructed might appear to be
important, the question of commutation of the various operators
naturally arises. In this section we will deal with this in some
detail. As for pairs of $\Omega$ operators, all possible combinations
commute [see Fig.\ref{fig:rules}b(I)-(IV)]:

\ba
\lefteqn{
[ \Omega^{|J|}_{i j w_1} , \Omega^{|J|}_{j k w_2} ] = 0 \: ; \:\:
[ \Omega^{|J|}_{i j w_1} , \Omega^{|J|}_{i k w_2} ] = 0 } \nonumber \\
&&
[ \Omega^{|J|}_{i j w_1} , \Omega^{|J|}_{k j w_2} ] = 0 \: ; \:\:
[ \Omega^{|J|}_{i j w_1} , \Omega^{|J|}_{k l w_2} ] = 0 .
\label{eq:Omega-commute}
\ea

\noindent We demonstrate the calculation required to prove the first
of these relations (we drop the normalization factors and set $J=1$
for notational simplicity):

\end{multicols}
%%\hrulefill\

\ba
\lefteqn{
\Omega_{jkw_2} \Omega_{ijw_1} |s_i,s_j,s_k,w_1,w_2\ra = } \nonumber \\
&&
\Omega_{jkw_2} \left[
-s_j \left( x^{s_i} |s_i,s_j,s_k,-s_j w_1, w_2 \ra -
w_1 x^{-s_i} |s_i,s_j,s_k,s_j w_1, w_2 \ra \right)
\right] =  \nonumber \\
&&
s_k s_j \left[ x^{s_j+s_i} |s_i,s_j,s_k,-s_j w_1, -s_k w_2 \ra - w_2 x^{-s_j+s_i} |s_i,s_j,s_k,-s_j w_1, s_k w_2 \ra
\right. + \nonumber \\
&&
\left. -w_1 x^{s_j-s_i} |s_i,s_j,s_k,s_j w_1, -s_k w_2 \ra + w_1 w_2
x^{-s_j-s_i} |s_i,s_j,s_k,s_j w_1, s_k w_2 \ra \right] , \nonumber
\ea

\noindent whereas on the other hand:

\ba
\lefteqn{
\Omega_{ijw_1} \Omega_{jkw_2} |s_i,s_j,s_k,w_1,w_2\ra = } \nonumber \\
&&
\Omega_{ijw_1} \left[
-s_k \left( x^{s_j} |s_i,s_j,s_k,w_1, -s_k w_2 \ra -
w_2 x^{-s_j} |s_i,s_j,s_k,w_1, s_k w_2 \ra \right) \right] =  \nonumber \\
&&
s_j s_k \left[
x^{s_i+s_j} |s_i,s_j,s_k,-s_j w_1, -s_k w_2 \ra -
w_1 x^{-s_i+s_j} |s_i,s_j,s_k,s_j w_1, -s_k w_2 \ra
\right. + \nonumber \\
&&
\left. -w_2 x^{s_i-s_j} |s_i,s_j,s_k,-s_j w_1, s_k w_2 \ra +
w_1 w_2 x^{-s_i-s_j} |s_i,s_j,s_k,s_j w_1, s_k w_2 \ra
\right] . \nonumber
\ea

\noindent It can be verified that the last lines in these two
calculations are identical, proving the first commutation
relation. Next we consider combinations of $S$ and $\Omega$. They
all commute except one:

\ba
[ S^J_{i j} , \Omega^{|J|}_{i k w} ] = 0 \: ; \:\:
[ S^J_{i j} , \Omega^{|J|}_{k i w} ] = 0 \: ; \:\:
[ S^J_{i j} , \Omega^{|J|}_{k l w} ] = 0 \: ; \:\:
[ S^J_{i j} , \Omega^{|J|}_{j k w} ] ,
\label{eq:S-Omega-commute}
\ea

\ba
[ S^J_{i j} , \Omega^{|J|}_{k j w} ] |s_i,s_j,s_k,w\ra = 
{1 \over {\sqrt{c_i c_k}}} x^{J s_i} \left( x^{J s_k} + w x^{-J s_k} \right)
|s_i,-s_j,s_k\ra \left[ |s_j w\ra - |-s_j w \ra \right]
\label{eq:S-Omega-dont-commute}
\ea

\noindent We demonstrate this, again taking $J=1$ and dropping 
normalization:

\ba
\lefteqn{
\Omega_{kjw} S_{ij} |s_i,s_j,s_k,w\ra = } \nonumber \\
&&
\Omega_{kjw} \left[
x^{-s_i} |s_i,s_j,s_k,w\ra + s_j x^{s_i} |s_i,-s_j,s_k,w\ra
\right] = \nonumber \\
&&
-s_j \left( x^{s_k-s_i} |s_i,s_j,s_k,-s_j w\ra -
w x^{-s_k-s_i} |s_i,s_j,s_k,s_k w\ra + 
s_j x^{s_k+s_i} |s_i,-s_j,s_k,s_j w\ra -
w s_j x^{-s_k+s_i} |s_i,-s_j,s_k,-s_j w\ra
\right) ,
\nonumber
\ea

\noindent whereas:

\ba
\lefteqn{
S_{ij} \Omega_{kjw} |s_i,s_j,s_k,w\ra = } \nonumber \\
&&
-s_j S_{ij} \left[
x^{s_k} |s_i,s_j,s_k,-s_j w\ra - w x^{-s_k} |s_i,s_j,s_k,s_j w\ra
\right] = \nonumber \\
&&
-s_j \left[
x^{-s_i+s_k} |s_i,s_j,s_k,-s_j w\ra +
s_j x^{s_i+s_k} |s_i,-s_j,s_k,-s_j w\ra -
w x^{-s_i-s_k} |s_i,s_j,s_k,s_j w\ra - w s_j x^{s_i-s_k} |s_i,-s_j,s_k,s_j w\ra
\right] ,
\nonumber
\ea

%%\hrulefill\
\begin{multicols}{2}

\noindent which is different from the previous result in the second
and fourth terms. The reason that the relation 
in Eq.(\ref{eq:S-Omega-dont-commute}) does not commute is that only
$\Omega_{kjw} S_{ij}$ produces the right result. In the opposite order
the same problem arises as when one tries to close a loop with $S$'s
only. This relation is depicted graphically in
Fig.\ref{fig:rules}(d).

Given the commutation relations in Eqs.(\ref{eq:Omega-commute}) and
(\ref{eq:S-Omega-commute}) one is essentially free to connect the
isolated plaquettes in any order, as we prove next.

\subsubsection{Proof of the Algorithm}
\label{2D-proof}

\begin{figure}
\hspace{2em}
\epsfysize=8.cm
\epsfxsize=8.cm
\epsffile{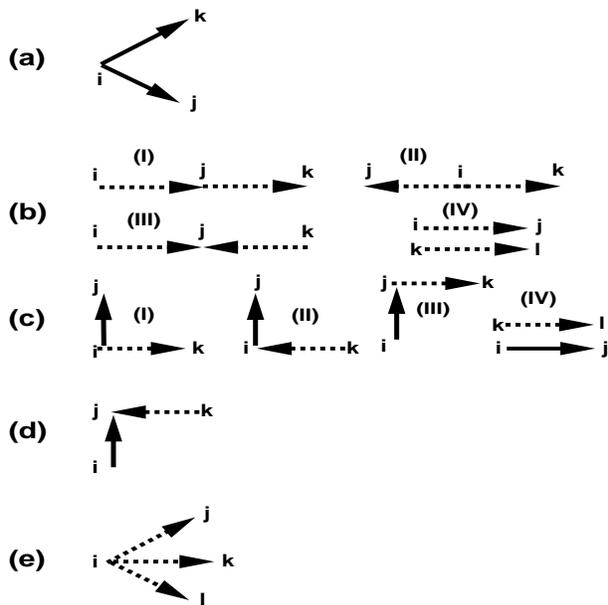}
\caption{All possible commutation relations of $S$ (full arrows) and
$\Omega$ (dashed arrows). (a) $[S_{ij},S_{ik}]=0$ (needed for the
Bethe lattice). (b) (I)-(VI) All combinations of $\Omega$ operators
commute. (c) The commuting combinations of $S$ and
$\Omega$: (I) $[S_{ij},\Omega_{ikw}]=0$, (II)
$[S_{ij},\Omega_{kiw}]=0$, (III) $[S_{ij},\Omega_{jkw}]=0$, (VI) $[S_{ij},\Omega_{klw}]=0$. (d) The
non-commuting combination of $S$ and $\Omega$. (e) Additional
commuting combinations needed in 4D and higher.}
\label{fig:rules}
\end{figure}

For simplicity we consider the case of a square lattice, where one has
prepared a set of 2$\times$2 plaquettes and placed them with equal
spacing on an $N$ by $M$ lattice (Fig.\ref{fig:2D}). Denote the Hamiltonian for
this system by $H^{\bf J}_{0}$. Clearly the maximum density
of non-overlapping prefabricated 2$\times$2 plaquettes which can be
achieved is 1/4. This can be increased by using larger plaquettes, at
the price of increasing complexity in their fabrication. The problem
is now to connect plaquettes, and from Corollary \ref{cor} this can be done with $\Omega$
operators which connect two {\em occupied} lattice points. The
geometries which may arise in connecting plaquettes are 
summarized in Fig.\ref{fig:rules}(b-d). The commutation relations of
the previous section show that the only problem can arise in the
geometry depicted in Fig.\ref{fig:rules}(d). However, as long as $S$
is applied {\em before} $\Omega$, the outcome is a correct
superposition provided $k$ Fig.\ref{fig:rules}(d) is the index of an
occupied site. The commutation relations assure that in any order in
which the $\Omega$'s are applied the lattice is generated
correctly. We may thus assume that some arbitrary sequence of $\Omega$'s has
been applied. We assume further that workbits corresponding to new bonds are
measured after application of $\Omega$, so that they are no longer in a
superposition and a bond with random sign has become integrated into
the lattice. It will then suffice to prove that introducing a bond
at an arbitrary location in the existing lattice produces the correct
superposition. 
Let us proceed by induction, and assume that some partial set of all bonds
has been closed by the algorithm. These bonds can be either
horizontal or vertical, and for definiteness we will assume that they
are always closed rightwards or upwards. Let us denote the set of
$K$ vertical bonds by $\{(n,m)\}_{K}$ and the $K'$ horizontal bonds by
$\{(n',m')\}_{K'}$. The Hamiltonian for this set is:

\ba
\lefteqn{
H^{\bf J}_{(K,K')} = H^{\bf J}_{0} + \sum_{\{(n,m)\}_{K}}
J_{(n,m)} s_{n,m} s_{n,m+1}
} \nonumber \\
&&
+ \sum_{\{(n',m')\}_{K'}} J'_{(n',m')} s_{n',m'} s_{n'+1,m'} ,
\label{eq:Hnm_K}
\ea

\noindent with a corresponding operator:

\ba
\lefteqn{
T^{\bf J}_{(K,K')} \equiv \prod_{\{(n,m)\}_K}
\left[ M_{w_{n,m}} \Omega^{|J_{(n,m)}|}_{(n,m)(n,m+1)w_{n,m}} \right]
\times
} \nonumber \\
&&
\prod_{\{(n',m')\}_{K'}}
\left[ M_{w_{n',m'}}
\Omega^{|J'_{(n',m')}|}_{(n',m')(n'+1,m')w_{n',m'}} \right]
T^{\bf J}_{0} .
\label{eq:T(nm)}
\ea

\noindent Here $M_{w_{n,m}}$ represents the measurement of workbit
$|w_{n,m}\ra$. The notation $\prod_{\{(n,m)\}_K}$ implies a product over
all members of the set $\{(n,m)\}_K$, in any order, and similarly for
the horizontal bonds (guaranteed correct due to the commutation
relations). Since the measurements have taken place, the workbits are
no longer in a superposition after $T^{\bf J}_{(K,K')}$ has been
applied. Therefore the induction hypothesis is:

\ba
\lefteqn{ T^{\bf J}_{(K,K')} |\{-\}_{NM}\ra |\{-\}_{K+K'}\ra = }
\nonumber \\
&& \sum_{y=0}^{2^{N M}-1} {1\over \sqrt{Z^{\bf J}_{(K,K')}}}
e^{-\beta H^{\bf J}_{(K,K')}\left[\{s\}_{y}\right]/2} |\{s\}_{y}\ra .
\label{eq:T(nm)-induction}
\ea

\noindent To see this it is helpful to consider for a moment the
situation that would arise without the intermediate measurements: for
every new (horizontal or vertical) bond $|J_{n,m}|$ one closes, one
needs to introduce a new workbit $w_{n,m}$. A total of $K+K'$ bonds
thus requires a workbit-register $|\{-\}_{K+K'}\ra$ as above. After
$|J_{n,m}|$ is closed, the workbit is in a superposition of states
corresponding to ${\rm sign}(J_{n,m}) = \pm 1$. This superposition is
destroyed after the measurements, leaving the spin-qubits'
superposition intact. This is the content of Eq.(\ref{eq:T(nm)-induction}).

The induction proof now requires showing that closing an additional bond, say
$|J_{p,q}|$, produces the correct superposition
\cite{QC:comment5}. The calculation is essentially identical to that
of Eq.(\ref{eq:calT}). It is easily checked that (i) this calculation
is insensitive to whether $|J_{p,q}|$ corresponds to a horizontal or
vertical bond, and (ii) the appearance of a double index for every spin
does not make any difference. There is thus no need to repeat the
calculation of Eq.(\ref{eq:calT}), and we conclude the 2D algorithm to
be proved.

It should be noted that every intermediate stage in the construction
of the full 2D lattice corresponds to a {\em diluted} 2D lattice,
with bonds specified deterministically by $H^{\bf J}_{0}$, and
other bonds being either ferro- or antiferromagnetic depending on the
result of measuring the workbits. Thus it follows that the
algorithm can be used to simulate diluted Ising models as well, with
the same restrictions applying to the generality of the class of these models as
specified above, i.e., the size of frustrated plaquettes must be finite.

\subsubsection{Control of Bond Sign}
\label{complexity}

One may wonder whether the lack of control over the bond sign is
special to the essentially 1D situation of closing an isolated
plaquette. In fact the same problem arises whenever a bond is closed
using $\Omega$, as we argue next. Let us suppose that the algorithm has correctly produced
the thermodynamic weights of the Hamiltonian $H_0 = -\sum_{\la i,j
\ra} J_{ij}s_i s_j$, describing {\em part} of the full 2D problem, and
excluding in particular the bond $J_{nm}$. Associated with $H_0$ is a
partition function $Z_0 = \sum_{\{s\}} \exp(-\beta
H_0[{\{s\}}])$. Let $p_{nm} \equiv {\rm Pr}[{\rm
sign}(J_{nm})=1]$ be the probability of a ferromagnetic bond, $q_{nm}
\equiv {\rm Pr}[{\rm sign}(J_{nm})=-1]$ that of an
antiferromagnetic bond. When the new bond $J_{nm}$ is included, we have for
the ratio of these probabilities:

\be
{{p_{nm}} \over {q_{nm}}} =
{{\sum_{\{s\}} e^{-\beta H_0[{\{s\}}]} e^{\beta |J_{nm}|s_n s_m}} \over
{{\sum_{\{s\}} e^{-\beta H_0[{\{s\}}]} e^{-\beta |J_{nm}|s_n s_m}}}} .
\label{eq:Pr}
\ee

\noindent This expression can be bounded from above and below by
replacing $s_n s_m$ in $\exp(\pm \beta |J_{nm}|s_n s_m)$ by $+1$ or
$-1$:

\ba
\lefteqn{ e^{-|J_{nm}|} \sum_{\{s\}} e^{-\beta H_0[{\{s\}}]} \leq }
\nonumber \\
&&
\sum_{\{s\}} e^{-\beta H_0[{\{s\}}]} e^{-\beta |J_{nm}|s_n s_m} ,
\sum_{\{s\}} e^{-\beta H_0[{\{s\}}]} e^{\beta |J_{nm}|s_n s_m} \nonumber \\
&&
\leq e^{\beta |J_{nm}|} \sum_{\{s\}} e^{-\beta H_0[{\{s\}}]} .
\ea

\noindent Inserting this into Eq.(\ref{eq:Pr}) yields:

\be
x^{-4|J_{nm}|} \leq
{{p_{nm}} \over {q_{nm}}}
\leq x^{4|J_{nm}|} .
\ee

\noindent This result is very similar to that obtained in the 1D case,
Eq.(\ref{eq:weight:2}), and although we cannot perform an explicit
calculation here, the implications are likely to be the same. Namely,
that the upper bound is approached as $T\!\rightarrow\!0$ so that a
frustrated plaquette becomes $\exp(2\beta J)$ less likely than an
unfrustrated one. The main difference compared to the 1D case is that
now closing a single bond $J_{nm}$ may correspond to closing several
plaquettes at once, which can only amplify the effect. Thus the
issue of control of the bond sign is even more problematic in 2D and
above. Nevertheless, since isolated plaquettes can be
``prefabricated'', the algorithm covers an exponentially large class
of Ising models, as argued above.

\subsection{3D Ising Model}
\label{3D}

Three-dimensional Ising models are notoriously difficult,
both analytically and computationally. In particular, whereas a large
number of 2D models have yielded to analyses \cite{Baxter:book}, only
very few 3D systems have been solved analytically
\cite{Baxter:6,Baxter:7,Suzuki,Huang}. Even so, all these models
suffer from various shortcomings, such as having negative Boltzmann
weights \cite{Baxter:6,Baxter:7} or being essentially 2D
\cite{Suzuki,Huang}. Computationally, the plagues of 2D are of course
multiplied in 3D, making our knowledge of 3D systems
limited. Moreover, there are good reasons to suspect that calculations
in 3D are {\em fundamentally} harder than in 2D. For example, some 3D
problems, such as finding the spin glass ground state, are known to be
NP-hard \cite{Barahona,Parisi86,Anderson86}. This is related to the existence of a freezing transition
at $T_c >0$ in 3D, in contrast to 2D, where $T_c=0$
\cite{Morgenstern}. Physically, this means that for $T<T_c$ the system
can get stuck in a local minimum in 3D (ergodicity breaking) and never
reach the ground state. Thus it is extremely important to develop
efficient algorithms for 3D systems. Let us now describe the extension
to 3D of the 2D algorithm presented in the previous section.

\subsubsection{Algorithm for 3D and Higher Dimensions}
The algorithm for the 3D case is a natural extension of that for 2D
and presents no essentially new problems. One can either prepare 2D
plaquettes and stack planes connected by $\Omega$ operators, or
prefabricate 3D cubes (using the methods of Sec.\ref{interference})
and connect those with $\Omega$'s. In both cases Corollary \ref{cor}
and the commutation properties of the $\Omega$ operators guarantee
that the correct superposition is produced, and the order in which
plaquettes or cubes are hooked up does not matter. The only new feature
is that more than two bonds can now emanate from the same site (for a
cubic lattice). However, these present no problem since the algorithm
is invariant to the order of creation of such bonds, as they all
commute in pairs. In other words, the induction
proof for the 2D case holds here as well, and the 3D algorithm is
proved. Clearly, also the complexity remains the same, namely $O(N)$
as long as random bonds between plaquettes are allowed. Moreover, the
same argument holds for any dimensional Ising system.

\section{Including a Magnetic Field}
\label{field}

In this final section we show how the present algorithm can be
extended in order to deal with the Ising model in the presence of a
magnetic field. The idea is to generalize the basic $R$ and $S$
operators introduced in the 1D case. Let:

\be
R^{\Delta_i}_i |s_i\ra = 
{1\over \sqrt{c_+^{0,\Delta_i}}}
\left( x^{s_i\Delta_i} |-\ra - s_i x^{-s_i\Delta_i} |+\ra \right) ,
\label{eq:R-Delta}
\ee

\noindent and

\ba
\lefteqn{
S^{J_{i},\Delta_j} | s_i,s_j \ra = } \nonumber \\
&&
{1 \over \sqrt{c^{J_{i},\Delta_{j}}_{s_i}}} \left(
x^{-(J_{i} s_i + \Delta_j)} | s_i,s_j \ra +
s_j x^{J_{i} s_i + \Delta_j} | s_i,-s_j \ra
\right) ,
\label{eq:S^J,Delta_i}
\ea

\noindent where:

\be
c^{J_{i},\Delta_{j}}_{s_i} = 2\cosh[\beta (J_{i} + s_i \Delta_j)] .
\label{eq:c}
\ee

\noindent It is easily checked that $R^{\Delta_i}_i$ and
$S^{J_{i},\Delta_j}$ are unitary. The loop-closing operator $\Omega$
can be generalized accordingly:

\ba
\lefteqn{ \Omega^{|J_i|,|\Delta_j|}_{i,j,w}|s_i,s_j,w\ra \equiv
X_{j,w}S^{|J_i|,|\Delta_j|}_{i,w}|s_i,s_j,w\ra = } \nonumber \\
&&
{-s_j \over \sqrt{c_{s_i}^{|J_i|,|\Delta_j|}}} \left( x^{|J_i| s_i +
|\Delta_j|}|s_i,s_j,-s_j w\ra - \right. \nonumber \\
&&
\left. w x^{-(|J_i| s_i + |\Delta_j|)}|s_i,s_j,s_j w\ra \right) .
\label{eq:Omega_ijw:2}
\ea

\subsection{Open Chain Case}

In order to introduce an arbitrary magnetic field on every spin in an
open chain geometry, consider the effect of applying $R$ and $S$ on a
two-qubit register:

\ba
\lefteqn{ S_{12}^{J_1,\Delta_2} R_1^{\Delta_1} |--\ra = } \nonumber \\
&&
{1 \over \sqrt{c_+^{0,\Delta_1}}}
\sum_{s_1,s_2} {-s_2 \over \sqrt{c_{s_1}^{J_1,\Delta_2}}}
x^{J_1 s_1 s_2 + \Delta_1 s_1 + \Delta_2 s_2} |s_1,s_2\ra ,
\label{eq:x1}
\ea

\noindent omitting some intermediate lines of by now
familiar algebra. The coupling of $\Delta_i$ to the $s_i$ in the
exponent is like that of a magnetic field. However, the normalization factor
also depends on $s_1$, so it needs to be considered as well:

\ba
\lefteqn{
\log c^{J,\Delta}_s = } \nonumber \\
&&
{1 \over 2} \left[
\log\left( e^{\beta(J + \Delta)} + e^{-\beta(J + \Delta)} \right) + \right.
\nonumber \\
&&
\left.
\log\left( e^{\beta(J - \Delta)} + e^{-\beta(J - \Delta)} \right) \right] +
\nonumber \\
&&
{s \over 2} \left[
\log\left( e^{\beta(J + \Delta)} + e^{-\beta(J + \Delta)} \right) - \right.
\nonumber \\
&&
\left.
\log\left( e^{\beta(J - \Delta)} + e^{-\beta(J - \Delta)} \right) \right] =
\nonumber \\
&&
{1 \over 2} \log\left(4\cosh[\beta(J+\Delta)]\cosh[\beta(J-\Delta)] \right) +
\nonumber \\
&&
{s \over 2} \log\left( {\cosh[\beta(J+\Delta)] \over
\cosh[\beta(J-\Delta)]} \right) \nonumber ,
\ea

\noindent so that: 

\ba
\lefteqn{
{1 \over \sqrt{c_{s}^{J,\Delta}}} =
\exp\left(-{1 \over 2} \log c_{s}^{J,\Delta}\right) = } \nonumber \\
&&
\left[ c_-^{J,\Delta} c_+^{J,\Delta} \right]^{-1/4} x^{-{1 \over
2\beta} s \log\left(c_+^{J,\Delta}/c_-^{J,\Delta}\right)} .
\label{eq:x2}
\ea

\noindent Collecting the exponents of $x$ in Eqs.(\ref{eq:x1}) and
(\ref{eq:x2}) we find for the two-qubit Hamiltonian:

\be
H^o_2 =
-J_1 s_1 s_2 -
\left[ \Delta_1 -
{1 \over 2\beta} \log\left( {c_+^{J_1,\Delta_{1+1}} \over
c_-^{J_1,\Delta_{1+1}} } \right) \right] s_1 - \Delta_2 s_2 .
\label{eq:H2}
\ee

\noindent This suggests that in general the magnetic field on spin $i$
simulated by the algorithm takes the form:

\be
h_i = \Delta_i - {1 \over 2\beta} \log\left( {c_+^{J_i,\Delta_{i+1}}
\over c_-^{J_i,\Delta_{i+1}}} \right) ,
\label{eq:h_i}
\ee

\noindent whence the $N$-spin Hamiltonian for an {\em open} 1D chain becomes:

\be
H^o_N = -\sum_{i=1}^{N-1} J_i s_i s_{i+1} -\sum_{i=1}^{N-1} h_i s_i -
\Delta_N s_N .
\label{eq:H_No}
\ee

\noindent Note that $h_i$ can take any value for a given choice of
finite $J_i$ and $\beta$, by tuning the parameters $\Delta_i, \Delta_{i+1}$. To
prove that the algorithm simulates an Ising model with the Hamiltonian
of Eq.(\ref{eq:H_No}) we proceed by induction. Assuming:

\ba
\lefteqn{
\prod_{i=N-1}^1 S_{i,i+1}^{J_i,\Delta_{i+1}} R_1^{\Delta_1} |\{-\}_N\ra =
} \nonumber \\
&&
\omega_N \sum_{y=0}^{2^N-1} \phi_N x^{-H^o_N[\{s\}_y]} |\{s\}_y\ra \:;
\nonumber \\
&&
\omega_N \equiv {1 \over \sqrt{c_+^{0,\Delta_1}}} \prod_{i=1}^{N-1}
\left[ c_-^{J_i,\Delta_{i+1}} c_+^{J_i,\Delta_{i+1}} \right]^{-1/4} ,
\label{eq:x3}
\ea

\noindent [where $\phi_N$ is defined in Eq.(\ref{eq:phi_N})] consider:

\ba
\lefteqn{
\prod_{i=N}^1 S_{i,i+1}^{J_i,\Delta_{i+1}} R_1^{\Delta_1} |\{-\}_{N+1}\ra =
} \nonumber \\
&&
\omega_N \sum_{y=0}^{2^N-1} \phi_N x^{-H^o_N[\{s\}_y]}
|\{s_1,...s_{N-1}\}_y\ra S_{N,N+1}^{J_N,\Delta_{N+1}} |s_N,-\ra =
\nonumber \\
&&
\omega_{N+1} \sum_{y=0}^{2^{N+1}-1} \phi_{N+1} x^{-H^o_{N+1}[\{s\}_y]}
|\{s\}_y\ra ,
\ea

\noindent where we used Eqs.(\ref{eq:S^J,Delta_i}) and
(\ref{eq:x2})-(\ref{eq:h_i}). This proves the algorithm for the open
chain case.

\subsection{Closed Chain Case}
The algorithm for the closed chain geometry takes a somewhat different
form. Instead of applying $R_1^{\Delta_1}$ one applies an ordinary
$\pi/2$ rotation on the first qubit, and closes the loop with
$\Omega^{|J_N|,|\Delta_1|}_{N1w}$. This results, as usual, in a
superposition of ferro- and antiferromagnetic last bonds, but also of
positive and negative fields on $s_1$. To see this, consider:

\ba
\lefteqn{
|\psi\ra = \prod_{i=N-1}^1
S_{i,i+1}^{J_i,\Delta_{i+1}} R_1 |\{-\}_{N}\ra = }
\nonumber \\
&&
\prod_{i=1}^{N-1}
\left[ c_-^{J_i,\Delta_{i+1}} c_+^{J_i,\Delta_{i+1}} \right]^{-1/4}
\sum_{y=0}^{2^N-1} \phi_N x^{-H'_N[\{s\}_y]} |\{s\}_y\ra
\ea

\noindent where

\ba
\lefteqn{ H'_N = -\sum_{i=1}^{N-1} J_i s_i s_{i+1} }
\nonumber \\
&&
- {\sum_{i=2}^{N-1} h_i s_i} +
(1/2\beta)\log(c_+^{J_1,\Delta_2}/c_i^{J_1,\Delta_2}) - {\Delta_N
s_N}. \nonumber
\ea

\noindent When we now introduce a workbit and apply $\Omega$, we find,
after a calculation similar to Eq.(\ref{eq:calT}):

\ba
\Omega^{|J_N|,|\Delta_1|}_{N1w} |\psi\ra |w=-\ra = \prod_{i=1}^{N}
\left[ c_-^{J_i,\Delta_{i+1}} c_+^{J_i,\Delta_{i+1}} \right]^{-1/4}
\times
\nonumber \\
\sum_{w=\pm 1} \sum_{y=0}^{2^N-1} \Phi_N x^{-H^c_N[\{s\}_y]}
|\{s\}_y\ra |w\ra , \nonumber \\
\label{eq:closed}
\ea

where $\Delta_{N+1} \equiv \Delta_1$,

\be
H^c_N = -\sum_{i=1}^{N} J_i s_i s_{i+1} -\sum_{i=1}^{N} h_i s_i .
\label{eq:H_Nc}
\ee

\noindent $H^c_N$ is seen
to be the correct Hamiltonian for a closed loop in the presence of the
local fields $h_i$. The sum over
$w=\pm 1$ in Eq.(\ref{eq:closed}) is such that $w\!=\!{\rm
sign}J_N\!=\!{\rm sign}\Delta_1$, so that indeed, the algorithm
produces a superposition over bond and field signs. Selecting a
particular sign can be done with the interference method of
Sec.\ref{interference}, and the plaquette thus generated can be
integrated into a higher dimensional lattice. A bond connecting
plaquettes should not have to include a field term, since it presumable
connects spins which already have a field on them from the plaquette
fabrication stage. Thus the situation in terms of controlling the
introduction of a magnetic field is better than that of the bonds:
arbitrary fields can be generated by the algorithm with full control
over the field at every lattice point. It is interesting to point out
in this context that it is known that the 2D {\em fully
antiferromagnetic} Ising model with equal interactions, in the
presence of a {\em constant} magnetic field is an NP-hard problem
\cite{Barahona}. 

\section{Conclusions and Outlook}
\label{conclusions}

To conclude, we have introduced a new approach for the numerical study
of statistical mechanics of Ising spin systems on quantum computers.
The approach consists of an algorithm which allows one to construct a
superposition of qubit states, such that each state uniquely codes for
a single configuration of Ising spins. Some stages of the algorithm,
such as the construction of the open 1D chain, are equivalent to the
Markov process used in the transfer matrix formalism. 
Closing loops can be
done by repeated measurements, requiring $\sim
\exp(4\beta J)$ attempts per frustrated plaquette. 
A crucial stage in the present algorithm,
which leads to a polynomial increase in efficiency over 
algorithms which are based on measurements alone, 
is the use of an interference transformation. This
transformation eliminates part of the superposition and thus
determines whether the given plaquette will close in a frustrated or
unfrustrated configuration. This is done in one step compared to $\sim
\exp(4\beta J)$ attempts per frustrated plaquette 
in algorithms based on measurements alone. 
A central feature of the
algorithm is that the quantum probability of each state in the superposition is
exactly equal to the thermodynamic weight of the corresponding
configuration. When a measurement is performed, it causes the
superposition to collapse into a single state. The probabilities of
measuring states are ordered by the energies of the corresponding spin
configurations, with the ground state having the highest
probability. Therefore, statistical averages needed for calculations
of thermodynamic quantities obtained from the partition function, are
approximated in the fastest converging order in the number of
measurements. Unlike Monte Carlo simulations on a classical computer,
consecutive measurements on a quantum computer are totally
uncorrelated. 

The algorithm applies to a
large class of Ising systems, including 
partially frustrated models. A magnetic field
can be incorporated as well without increase in the complexity, which
is linear in the number of spins and bonds. The main problem of the
algorithm is the limited control it offers in the construction of a
{\em specific} realization of bonds on the Ising lattice. An attempt
to control {\em all} the bonds (and not only the prefabricated ones)
by repeating measurements may result in an exponential slowdown in performance as the
temperature is lowered, and for this reason the algorithm fails to address the
question of whether 
polynomial time (P) equals NP on a quantum computer, in the context of
finding the spin glass ground state. 
In summary, this paper provides tools for the simulation of Ising
spin systems on a quantum computer, as efficiently as the best
classical algorithms. 
Work employing these tools to achieve speedup over classical algorithms
is in progress.

\section*{Acknowledgements}
We would like to thank 
Dorit Aharonov, Nir Barnea, Eli Biham, Benny Gerber,
Guy Shinar, Haim Sompolinsky and Umesh Vazirani for
useful discussions and
Robert Griffiths, Michael Ben-Or and David
DiVincenzo for comments on an earlier version of the manuscript.

\end{multicols}

\newpage

\section*{Figures}

%%\hrulefill\

%%\hrulefill\

%%\hrulefill\
\newpage
%%\hrulefill\

%%\hrulefill\


\begin{thebibliography}{10}
\bibitem{QC:comment2}
{$S^{J_i}$ should be understood as ``S-plus-J'' or ``S-minus-J'' -- not ``$S$
  to the power $J_i$''.}
\bibitem{QC:comment3}
{Although one should keep in mind that the number of bits required for a
  desired accuracy of the computation must remain $O[\log_{10}(N)]$
  \cite{Shor:94,Bennett:2}.}
\bibitem{QC:comment4}
{This operator is still ``valid'', since it can be decomposed into a product of
  two-qubit operators.}
\bibitem{QC:inequality}
Suppose, without loss of generality, that $a_1/b_1 = \max_i^n (a_i/b_i)$. Then:
  $\sum_i^n a_i/\sum_i^n b_i = (a_1/b_1)(b_1/\sum_i^n b_i) + \sum_{j=2}^n
  (a_j/b_j)(b_j/\sum_{i=1}^n b_i) \leq (a_1/b_1)\sum_{i=1}^n (b_i/\sum_{i=1}^n
  b_i) = a_1/b_1$. For the other direction, take $a_1/b_1 = \min_i^n
  (a_i/b_i)$. This proves Eq.(\ref{eq:ineq}).
\bibitem{QC:comment5}
{Technically this is an induction on a finite set, since the lattice is finite.
  But this is not a problem, since once the entire lattice is covered the
  operation of closing a new bond is undefined.}
\end{thebibliography}
\end{document}